# Sensor Compendium
## - A Snowmass Whitepaper -

M. Artuso[1], M. Battaglia[2], G. Bolla[3], D. Bortoletto[3], B. Cabrera[16], J.E. Carlstrom[4], C.L. Chang[4], W. Cooper[9], C. Da Via[18], M. Demarteau[19], J. Fast[11], H. Frisch[15], M. Garcia-Sciveres[6], S. Golwala[14], C. Haber[6], J. Hall[11], E. Hoppe[11], K.D. Irwin[16], H. Kagan[7], C. Kenney[12], A.T. Lee[6], D. Lynn[10], J. Orrell[11], M. Pyle[17], R. Rusack[13], H. Sadrozinski[2], M.C. Sanchez[5], A. Seiden[2], W. Trischuk[8], J. Va'vra[12], M. Wetstein[4], R-Y. Zhu[14]

[1] *Syracuse University*
[2] *University of California Santa Cruz*
[3] *Purdue University*
[4] *University of Chicago/Argonne National Laboratory*
[5] *Iowa State University/Argonne National Laboratory*
[6] *Lawrence Berkeley National Laboratory*
[7] *Ohio State University*
[8] *University of Toronto*
[9] *Fermi National Accelerator Laboratory*
[10] *Brookhaven National Laboratory*
[11] *Pacific Northwest National Laboratory*
[12] *Stanford Linear Accelerator Laboratory*
[13] *University of Minnesota*
[14] *California Institute of Technology*
[15] *University of Chicago*
[16] *Stanford University*
[17] *University of California, Berkeley*
[18] *The University of Manchester*
[19] *Argonne National Laboratory*

## Abstract

Sensors play a key role in detecting both charged particles and photons for all three frontiers in Particle Physics. The signals from an individual sensor that can be used include ionization deposited, phonons created, or light emitted from excitations of the material. The individual sensors are then typically arrayed for detection of individual particles or groups of particles. Mounting of new, ever higher performance experiments, often depend on advances in sensors in a range of performance characteristics. These performance metrics can include position resolution for passing particles, time resolution on particles impacting the sensor, and overall rate capabilities. In addition the feasible detector area and cost frequently provides a limit to what can be built and therefore is often another area where improvements are important. Finally, radiation tolerance is becoming a requirement in a broad array of devices. We present a status report on a broad category of sensors, including challenges for the future and work in progress to solve those challenges.

# Table of Contents

# Introduction: Sensors[1]

Sensors play a key role in detecting both charged particles and photons for all three frontiers in Particle Physics. The signals from an individual sensor that can be used include ionization deposited, phonons created, or light emitted from excitations of the material. The individual sensors are then typically arrayed for detection of individual particles or groups of particles.

Mounting of new, ever higher performance experiments, often depend on advances in sensors in a range of performance characteristics. These performance metrics can include position resolution for passing particles, time resolution on particles impacting the sensor, and overall rate capabilities. In addition the feasible detector area and cost frequently provides a limit to what can be built and therefore is often another area where improvements are important. Finally, radiation tolerance is becoming a requirement in a broad array of devices. The system performance also depends crucially on the electronics chain used for readout. In many cases issues such as mechanical support, cooling, cabling and voltage distribution are critical to good performance. Some specific findings and recommendations are given below.

Findings and Recommendations: Summary

1. Low mass, high radiation requirements imply thinner detectors with lower signal levels, thus implying low noise electronics, with zero-suppressed readout (because of the high channel count) implemented with discriminators featuring low and stable thresholds. This will require significant work on electronics design as well as use of the latest chip technologies. A large number of sensor choices are under investigation, which have different advantages in terms of radiation hardness, low mass, and high speed targeted to the different emphases of the three frontiers.

2. Fine pitch, connector-less, reliable electrical connections are key to achieve the low mass-fine pitch detector systems that are planned for future applications. This is an area in which much more work could be done.

3. The goal of minimizing the detector mass implies an integrated system design incorporating cooling and power distribution as primary design considerations. A coordinated effort to develop new materials, establish a data base of their properties, and implement the standardization of their quality assurance certification is key to multiple science frontiers and has a dramatic spin-off potential.

4. Cosmic frontier applications span a very large spectrum of technologies and techniques but often have in common the requirement for low and well known background levels. A community wide approach to radiopure materials and assay is likely to be very beneficial. Some of the novel sensors being developed are Transition Edge Sensors and Microwave Kinetic Induction Detectors, thick CCDs, and large Germanium detectors. Low activity photomultipliers are very important for experiments using noble liquid detectors.

5. A number of other photon detection and timing devices are very important for progress at several of the frontiers. This includes solid state photodetectors, and a number of devices with significantly improved timing capabilities.

6. Crystal material suitable for large scale calorimeters, capable of withstanding high fluences, and featuring faster time response fulfill a central role in intensity and energy frontier applications.

---

[1] Corresponding author(s) for this section: M. Artuso[1], A. Seiden[2]

# Summary:  Needs for the Energy Frontier[2]

A next application of sensors for the Energy Frontier can be expected to be for the vertex detectors and trackers for the upgrades of CMS and ATLAS.  The increased luminosity of the LHC poses severe challenges.  Since the luminosity of the LHC will increase, the number of interactions per crossing will also rise from 20 to at least 140 and possibly 200 every 25 ns.  In order to keep the occupancy below 1%, we must reduce each cell size, and increase the granularity, and therefore the number of readout channels. This trend requires the development of smaller detector elements and it introduces challenges since it necessitates many channels equipped with fast and thus power dissipating frontend electronics. Especially in the layers closest to the interaction region, the next generation of detectors must also reach high rate capabilities using integrated circuits and materials with increased tolerance to radiation dose and single event upsets.

Future trackers must function in this environment while maintaining or improving their performance. Pixel detectors are essential to seed tracking, select the primary vertex of hard-scatter events of interest, and mitigate the effect of the pileup.  To fully identify many physics signatures, trackers and vertex detectors must provide excellent position and momentum measurements for high-$p_T$ leptons. They also have to establish that leptons are in fact isolated. The material in the tracker and the vertex detectors affects the efficiency and resolution to detect isolated electrons, due to bremsstrahlung and nuclear interactions. Therefore it is critical to reduce this degradation by diminishing the material.  The performance of b-tagging algorithms based on displaced vertices also requires controlling the tails in the impact parameter measurement due to errors in pattern recognition and hadronic interactions within the active volume of the vertex detector. Therefore the upgrades strive to achieve a larger channel count while maintaining a careful control on the material inside the tracking. This has profound implication on the mechanical design of the support structure, the powering scheme, and the cooling.

For pixel vertex detectors we must develop sensors capable to survive doses of about $2\times10^{16}$ neq/cm$^2$ or provide easy replacement of the innermost pixel layers. The pixel readout chip must handle rates of about 1GHz/cm$^2$. This requires smaller, faster pixel chips, with increased functionality.  Both CMS and ATLAS are focusing on the development of new electronics using the 65 nm CMOS technology. Thin planar sensors [1] operated at high voltage (about 700V for 75 $\mu$m) show 60% charge collection efficiency after $10^{16}$ neq/cm$^2$.  Non-planar 3D sensors have been shown to be radiation hard up to doses of $5\times10^{15}$ neq/cm$^2$ and they will used in a new inner pixel layer, the ATLAS IBL [2].  R&D should occur to further increase their radiation hardness. The reduced bias voltage needed to operate 3D sensors is also of great interest since it could allow lower mass cooling systems.  Diamond remains an attractive option, even if concerns continue to exist regarding material availability and decreased signal collection after irradiation due to trapped charge forming a polizarization field.  Because of the smaller charge collected in diamond with respect to silicon, to take full advantage of diamond detectors new very low threshold readout chips must be developed. The DIAPIX development [3] is going in this direction and aims at using 3D electronics to achieve a threshold of about 1000e$^-$. For hybrid pixel detectors the smaller pixels are likely to have an area of about 2500$\mu$m$^2$. This enables increased resolution and better two-track separation. The pixel size and the material could be further reduced using monolithic pixels.  However, the development of radiation hard CMOS pixels for hadron colliders is at a very early stage.

The next generation trackers will cover an area of about 200 m$^2$, similar to the one installed in the current CMS detector. Also for the outer layers of the tracker, the increased luminosity requires the development of smaller size devices such as macro-pixel, strixels, or short strips, that will also augment

---

the achievable resolution. Pitch reduction will also improve the tracker performance at high $p_T$. The higher tracker granularity, especially the decrease in strip length down to centimetres or even millimetres, poses new challenges to sensor channel connectivity to the readout [4]. This can already be seen for example in the progress of the module design of the silicon micro-strips for the ATLAS upgraded tracker, where the sensitive strip length shrinks from 12 cm (obtained by daisy chaining two sensors) to 2.4 cm, forcing the placement of the hybrids directly over the surface of the sensor. This example of a connectivity problem is solved using "traditional" methods, but the anticipated higher channel density would certainly benefit from novel solutions such as multi-metal routing, dielectric-metal interposers, combined use of bump and wire bonding, through silicon via (TSV) structures on electronics and/or sensors. Double metal structures are in use since LEP and the Tevatron and also today in the LHCb VELO detector. This last example shows a complex routing of inner short strips to the outer edge of the sensors by means of second metal lines. This routing can lead to charge losses or an increase in noise. For short, such as strips (strixels) in the range of millimetres, bump bonding seems to be advantageous but a more elaborate routing to dedicated bonding areas may be beneficial to avoid covering the whole sensor with chips. This would reduce the overall mass and cost of the system. R&D is needed to better understand the performance and limitation of routing in multi metal layers, in-between channels using the first metal, and using deposited thin dielectric and metal layers (e.g. Multi-Chip Module deposited, MCM-D). MCM-D is a technology that can be applied to silicon strip modules and promises advantages in terms of integration complexity and material budget. It allows integrating the front-end hybrid, pitch adaptor and wire bonds on the silicon sensor by depositing dielectric and metal layers on silicon, where traces and vias are etched with high resolution to produce a PCB like structure. Initial positive results have been obtained with MCM-D layers but further work for improving the density of lines and contacts needs to be performed.

One of the key issues for strip sensors deployed in a future tracker is to withstand fluences of the order of $2\times10^{15}$ $n_{eq}$ $cm^{-2}$. While these specifications are not as stringent as for pixel detectors, R&D is still necessary in order to find cost effective radiation hard sensors. Dedicated R&D has been conducted both by ATLAS and CMS. The CERN-RD50 collaboration has also been instrumental in the study of the radiation issues for the HL-LHC. The measurements that have been conducted indicate that the n-on-p sensors match the requirements of the HL-LHC. Some advantages have been identified using oxygen enriched MCz material. CMS is also starting to evaluate the potential of deploying sensors fabricated on 8" wafers at Infineon.

An important requirement of future trackers motivated by the luminosity increase is the need for intelligent trackers providing fast momentum information to select events of interest. R&D is on-going on several options including reading out the tracker data only in regions of interest, and on detector data reduction by using the correlations between top and bottom sensors necessary to measure the track curvature, measuring the $p_T$ using associative memories.

The importance of reducing the material is especially critical for $e^+e^-$ colliders but, as stated above, it is also important for the HL-LHC. R&D for the ILC and CLIC pixel detectors has focused on using CMOS or SOI (Silicon on Insulator) technologies that combine particle detection and read-out electronics in the same device [5]. The adoption of these commercial technologies is especially interesting since it could also bring significant cost savings. The 90nm and 65nm CMOS technology are currently being evaluated as a promising solution to the integration of high-density circuitry and high-speed functionalities in a small pixel. Present CMOS sensors do not have a deep depletion volume and are using only the shallow depletion depths produced by the junction built in voltage as sensor volume. For hadron collider applications the most significant issues to be solved are radiation hardness and the possibility to deplete the sensitive layer. High-voltage CMOS pixels are currently being pursued for the LHC upgrades since the strong electric field in the fully depleted volume allows the fast charge collection needed for radiation tolerance. It also reduces the cluster size and improves two-track separation. The SOI process developed by KEK with Lapis Semiconductor using a high-resistivity handle wafer ensures large signals, low collection electrode capacitance, and low-power and therefore could provide radiation tolerant devices.

Strip or mini-strip sensors have larger cells size than typical pixel sensors, and thus higher capacitance and noise. Therefore sensors must collect a large signal from a significant depleted volume. To exploit the SOI or CMOS potential in these systems, novel design concepts have to be developed to combine small electronic feature size with high depletion voltage for the sensor bulk, without impacting on the performance of the electronics processed above the sensitive volume. R&D must be directed towards the application of these technologies to strip or mini-strip sensors to increase the efficiency of future trackers. Since a mini-strip cell size is relatively large this could yield the possibility of implementing additional intelligence in the available area such as large buffer sizes, sparsification, ADC, and trigger logic. This approach could help to reduce the material of tracker systems and solve the complex routing for the strixel/macro-pixel geometry. R&D in this area would be beneficial for the outer tracker and the inner pixel region of the LHC detectors and is relevant for future $e^+e^-$ colliders.

# Summary:  Needs for the Intensity Frontier[3]

Experiments planned for the "intensity frontier" encompass a broad spectrum of studies and technologies. Experiments optimized for neutrino detection will explore the neutrino mass hierarchy, search for CP violation and non-standard interactions, and elucidate the nature of neutrinos (Dirac or Majorana). In additions, large scale neutrino experiments will increase sensitivity to proton decay. Extremely rare muon and tau decay experiments will search for violation of charged lepton quantum numbers. Rare and CP-violating decays of bottom, charm, and strange particles will allow to probe new physics at very high mass scale and will be complementary to direct searches for new phenomena at the high energy frontier in our effort to unravel the mass scale and nature of new physics.

A key element in detector systems designed to study heavy quark and charged lepton decays is precision tracking and vertexing that relies on the most advanced Si based micropattern detectors. The subsystem in closer proximity to the interaction region involve pixel detectors, either monolithic (DEPFET) [6] or hybrid [7]. These subsystems are complemented by silicon microstrip detectors, where the sensor is generally based on a proven technology, but where the "material challenge" to minimize the non-active material crossed by the charged particles is of utmost importance.

Requirements on the front end electronics  processing information from Si devices for heavy quark experiments at the intensity frontier shows a general trend towards faster readout and synergy with the energy frontier.  For example, the BelleII readout electronics for the strip detector is based on the asic developed for CMS.  LHCb presents an interesting evolution of this concept, with fast analog readout and a processing block that allows to push the data out  of the detector in real time, and thus allows for a purely software based trigger selection.

The reconstruction and identification of photons, electrons and neutral pions is generally based on Ring Imaging Cherenkov detectors, which require R&D on faster photon detectors. An alternative based on picosecond time of flight technologies is gaining a lot of attention and may be a very attractive solution for K, p, and $\pi$ identification in certain momentum ranges.

Photosensors provide essential functionality for HEP experiments in a wide variety of contexts. Increasingly, the demands of HEP experiments are pushing beyond standard capabilities of conventional phototubes.  Photosensors are typically used in three broad categories of detector systems: (1) Large water Cherenkov tanks for full event reconstruction in neutrino and proton decay experiments, (2) Ring Imaging Cherenkov (RICH) and Direct Internally Reflected Cherenkov (DIRC) detectors used for particle identification in heavy flavor collider experiments, (3) and simple time-of-flight (TOF) systems used for vertex reconstruction and time-of-flight mass spectroscopy in colliders. The demands on these photosensors vary widely from one experiment to the next, but can be divided between the two main types of operational conditions: 1) Low-rate environments such as neutrino, double beta decay, and proton decay experiments. Under these conditions low cost coverage, improved resolution for event reconstruction, and single photon counting capabilities are most important for background rejection, signal selection efficiency, tracking, and momentum resolution. 2) High-rate environments typical of collider based experiments. Here photodetectors need to be robust over years of operation under intense radiation, and capable of handling high event rates per unit area per unit time without saturating. Readout electronics must be designed to handle the high rates and track multiplicities.

Common to all experimental categories is the need to instrument large areas economically and the need for improvements in the combined timing and imaging resolutions.

---

[3] Corresponding author(s) for this section:  M. Artuso[1], M. Wetstein[4], M.C. Sanchez[5]

# Summary: Needs for the Cosmic Frontier[4]

Experiments planned for the "cosmic frontier" span a tremendously vast spectrum of technologies and techniques from low-background experiments deep underground to vast air-shower detectors, to telescope and satellite-based survey instruments imaging the cosmos across a broad range of the electromagnetic spectrum. The primary aims of this frontier are the discovery and characterization of dark matter and elucidation of the nature of dark energy. In addition, experiments are probing the highest energy particles and photons in the universe.

The dark matter "generation 2" experiments are currently in the final phase of R&D with a down-select expected this winter to a set of 2-3 technologies to be fully developed for the next decade of searches. While there is a theoretical bias towards higher masses (several 10's or hundreds of GeV), there continue to be tantalizing excesses of events in several experiments that are consistent with a low-mass Weakly Interaction Massive Particle (WIMP) candidate. Since the nature of the dark matter remains a mystery, it is critical that the suite of experiments continues to probe the full spectrum of possible interactions – axions, spin-dependent and spin-independent couplings. Low radiometric background materials and assay techniques to verify that the materials used meet background goals is a common challenge across nearly all of these experiments. While the background requirements are not as strict as those for neutrinoless double-beta decay searches, they are nonetheless quite difficult to achieve. A community-wide approach to radiopure materials and assay is likely to be required as the generation 2 experiments are scaled up and generation 3 experiment concepts are developed.

Dark matter searches use a variety of approaches that utilize all three potential signals from a detector material – phonons, ionization and scintillation – in general trying to exploit two of the three to provide discrimination between WIMP recoil events and ionization events. Hence the sensor research program spans high-fidelity ionization spectrometers, high-performance phonon detectors such as the Transition Edge Sensor (TES) coupled to Superconducting Quantum Interference Devices (SQUIDs), and low-background PMTs operable at cryogenic temperatures for scintillation detection in liquid noble gas detectors. In addition, novel new approaches such as thick CCDs with extremely low thresholds are being pursued to extend the experimental reach to much lower WIMP masses.

Experiments observing the cosmic microwave background (CMB) are focusing R&D efforts on cryogenic Transition Edge Sensors (TES) and Microwave Kinetic Induction Detectors (MKID) that rely on superconducting phase transitions to measure minute temperature rises due to interactions of the CMB photons with the sensors. Scaling up to large arrays is a significant challenge for next-generation experiments.

Finally, there are the large air-shower arrays that will benefit from low-cost, large area photosensors that are under development for many applications across all of the high-energy physics frontiers.

---

[4] Corresponding author(s) for this section: J. Fast[11]

# Monolithic Pixel Sensors[5]

The standard figure of merit for vertex tracking accuracy in collider experiments is the accuracy on the impact parameter, $\sigma_{IP}$, defined as the distance of closest approach of the particle track to the colliding beam position. This can be parametrised as: $\sigma_{IP}$ = a + b/p $\sin^k \theta$, where p is the particle momentum, $\theta$ the track polar angle and k = 3/2 for the R-$\phi$ and 5/2 for the z projection. The identification of hadronic jets originating from heavy quarks is best achieved by a topological reconstruction of the displaced vertex structure and the kinematics associated to their decays. The ability in reconstructing these sequences of primary, secondary and tertiary vertices depends on the impact parameter resolution. This is related to the single point resolution of the sensors and their geometry. The figure of merit for the resolution is the distribution of the impact parameter values for charged particles originating from *b* hadron decays. It is useful to recall that the impact parameter does not depend, to a first approximation, on the *b* hadron energy, since the angle of production of the track in the lab frame is proportional to $(\beta\gamma)^{-1}$, which compensates the $\beta\gamma$ term in the *b* hadron decay length. The average impact parameter value is ~220$\mu$m, with a 0.15 (0.24) fraction of tracks having impact parameter below 15 (35) $\mu$m. The asymptotic resolution that appears in $\sigma_{IP}$ is typically proportional to the sensor point resolution, the tracking lever arm and, to a lesser extent, the radial position, R, of the first layer. The multiple scattering coefficient, b, scales with sqrt(d) R/p , where d is essentially the material thickness of the beam pipe and the first layer. The tracking lever arm is constrained by detector volume, number of electronics channels and cost. The minimum detector radius is typically defined by radiation or particle density conditions from particle fluence and beam-induced backgrounds and it is accelerator dependant. This was 63mm at LEP and 25mm at SLC, is 30-50mm at the LHC, with plans to further reduce it, and should be 15-30mm at future lepton colliders.

The two main drivers for the sensor pixel pitch, P, are single point resolution and two-track separation, both scaling as P, and local occupancy, scaling as the pixel area (i.e. proportional to $P^2$ for square pixels) times the readout time. As a result of this, starting with the current LHC detectors, fast time-stamping, required by occupancy, has become part of the overall optimization and the space-time granularity has to be considered as the appropriate parameter space for characterizing any future vertex detectors.

Monolithic pixel sensors, where the Si sensitive volume and the charge sensing and possibly part of the readout circuitry are implanted in the same Si wafer, have emerged at the end of the '90s in the framework of the R&D towards a future linear collider as the way forward towards thin Si sensors with very high granularity and small pixel capacitance, well beyond the technological capabilities for the hybrid pixel sensors then under development for application in the LHC experiments. This development was made possible by the progress in deep sub-micron CMOS processes and the availability of commercial processes with thick epitaxial layers aimed at sensors for optical imaging and consumer digital cameras, where CMOS image sensors have overtaken CCDs in terms of market share.

Monolithic pixels offer less mass than hybrid pixel detectors, lower pixel capacitance (and thus noise) and a single chip solution without bump bonding [8]. Combining particle detection and (at least part of) front-end electronics (FEE) in the same device is a prominent advantage for monolithic CMOS pixels, but this may also turn into a limitation since their manufacturing relies on industrial processes optimized for commercial products, which may depart substantially from those needed for charged particle detection. However, CMOS industry is evolving in a direction which allows CMOS pixels to progressively approach their real potential. Developments towards future vertex detector systems rely on the advances in the CMOS technology to cope with the demanding requirements of future collider experiments. To this purpose, many R&D paths are being pursued in the vertex detector community, with the US playing an important, although not yet a leading, role. The shrinking of the feature size of CMOS transistors is the most apparent outcome of the evolution of microelectronics. This is being exploited in the design of new pixels where advanced functions can be performed in the circuitry implemented in each pixel to provide the required signal-to-noise ratio and to handle the high data rate. These functions include amplification, filtering, calibration, discriminator threshold adjustment, data sparsification and time stamping.

---

[5] Corresponding author(s) for this section: M. Battaglia[2]

CMOS sensors for particle tracking were initially designed following the classical guidelines of the NMOS-only 3-transistor cell of image sensors to collect the charge generated in an almost field-free epitaxial layer. Now, thanks to CMOS scaling, more functions can be included in the pixel cell, including correlated double sampling and hit binary information from a discriminator. This leads itself to achieve zero suppression at the pixel level, which is a highly desirable feature given the large number of channels foreseen and the fast readout of the pixel matrix. New strategies are also being devised in terms of the needed readout architecture, introducing a higher degree of parallelism to speed up the transfer of information from pixels to the chip periphery. Another way to comply with a high hit rate is to adopt the solutions developed for hybrid pixel sensors. This requires that the classical continuous-time analog signal processing chain (charge-sensitive preamplifier, shaping filter) is integrated in the pixel cell, together with the digital blocks needed for a high-speed sparsified readout of the matrix [9].

The implementation of complex circuitry in the pixel cell requires a technique to insulate the transistors from the collection volume, to avoid parasitic charge collection and cross talk. This can be done using large deep N-well electrodes as sensing elements or by profiting from the thin oxide layer of the silicon-on-insulator (SOI) technology [10],[11] insulating the CMOS electronics to use full CMOS capabilities, without degradation of the charge collection efficiency.

These technologies make it possible to implement advanced readout architectures that can be tailored to different applications. However, special care needs to be taken to avoid analog-digital coupling. In the last few years, successful results have been reported for the 130nm CMOS technology node [12] and the 200nm SOI process [11]. Recently, the 180nm CMOS process featuring a nearly 20$\mu$m thick epitaxial layer with resistivity exceeding 1k$\Omega$ cm, 6 to7 metallization layers and additional P-wells offers the possibility to use both types of transistors inside the pixels, thus allowing for a substantial increase of the in-pixel micro-circuit complexity [13]. The 90nm and 65nm CMOS technology nodes are currently being evaluated as a promising solution to the integration of high density circuitry and high speed functionalities in a small pixel. High-voltage CMOS pixels [14] were pioneered several years ago with a larger feature size process [15] and their development is currently pursued on smaller feature size in the perspective of tracker upgrades at the LHC.

Besides device scaling, advanced CMOS technologies bring along other desirable features that pixel R&D projects are presently exploring. Metal interconnections are isolated by low-k dielectrics to reduce parasitic capacitance and time constants of transmission lines. This may be exploited by minimizing the electronic functions in the pixel cell and by connecting each pixel to signal processing blocks in the chip periphery. Thanks to the very high integration density of the CMOS process, this peripheral readout strategy may require just a small silicon area outside the pixel matrix itself. New strategies are also being devised in terms of the needed readout architecture, introducing a higher degree of parallelism to speed up the transfer of information from pixels to the chip periphery.

Monolithic CMOS and DEPFET technologies offer us sensors with thin active layers, pixel cells of O(10x10) $\mu m^2$ in size with a typical capacitance of O(1) fF and 10-40 $e^-$ ENC of noise and significant data processing capabilities implemented in-pixel or on-chip. Costs are low, as long as standard CMOS processes are used, and multi-project runs are widely available, also for several specialized processes.

The optimal pixel size for HEP applications is driven by practical considerations such as the readout-out rate and charge diffusion limit, but pixel pitches of 5μm and below are technically feasible and beneficial for specific imaging applications. An active thickness of 15-20μm is sufficient for achieving a detection efficiency >99% with CMOS pixels, even at operating temperatures in excess of 40°C. Back-thinning technologies, such as grinding and chemical etching now commonly available as commercial service, have been successfully applied to obtain 40-50 μm-thick CMOS sensors for the STAR HFT at RHIC and in the R&D for linear colliders and B-factories. Finally, the monolithic pixel technology is gaining ground in applications at the LHC, for example in the ALICE upgrade where the baseline design is based on monolithic pixels to obtain ladders with an overall material budget 0.3-0.4% $X_0$ / layer and the sensor contributing ~15%. DEPFET sensors can be thinned to less than 100 μm in the sensitive region, using wafer bonding technology, retaining a frame which ensures sufficient stiffness of the mechanical module. Thin DEPFETs are currently under development for BELLE-II, where the target thickness is 75 μm, and first results have already been obtained for prototypes successfully thinned to 50 μm. Beyond HEP there is a broad range of applications where monolithic pixel sensors have already been successfully tested, demonstrating superior performances compared to the detectors currently in use. They include X-ray imaging and spectroscopy from light sources and FELs to satellite astronomy, biological autoradiography and plasma diagnostics; low energy electron imaging for transmission electron microscopy and accelerator beam monitoring; optical biological imaging and low momentum particle detection in nuclear physics and real-time dose delivery assessment in hadrontherapy [16].

The use of a moderate- to high-resistivity thin substrate is the latest step in the process development exploited by monolithic pixel sensors. The depleted sensitive layer enhances both the amount of collected charge and the collection time also reducing the charge carriers spread, compared to the almost field-free epitaxial layer of the standard CMOS active pixel sensors, without compromising the sensor competitive thickness. Given that the reverse bias voltage, $V_d$, required to deplete a given thickness $d$ scales as $d^2 / \rho$, where $\rho$ is the Si resistivity, a modest $V_d$ of O(10 V) is sufficient to fully deplete a thin layer of high resistivity Si. Operating the detector fully depleted has advantages both in terms of signal charge and of collection time, thus improving the performance after irradiation by more than an order of magnitude, reducing at the same time the cluster size at the advantage of better two-track separation. Large signal-to-noise ratios are essential for near 100% efficiency and for implementing features such as time stamping. This can be achieved either with specialized CMOS processes featuring a resistivity in excess of 1kΩ cm, or with the SOI process on high resistivity handle wafer. Both approaches have been successfully exploited in monolithic pixel R&D ensuring signal-to-noise values of 30-40 from minimum ionizing particles at room temperature. The beam telescope developed within the EUDET EU project uses sensors based on high resistivity CMOS pixels thinned to 50 μm [17]. They provide a ~3μm single point resolution at a rate of ~$10^4$ frames/s over an active surface of about 2 cm$^2$ paved with 670k pixels and can cope with a flux of particles exceeding $10^6$/cm$^2$ s. The Silicon-On-Insulator process with a high-resistivity handle wafer, developed by an international consortium to which both LBNL and FNAL contribute significantly, offers further appealing opportunities by removing the inherent limitations of bulk CMOS processes. After the pioneering effort in a 3μm CMOS process [18], a commercial 0.2μm SOI process has been made available since a few years by a collaboration of KEK with Lapis Semiconductor Inc. (formerly OKI Semiconductors), Japan [19]. The high resistivity sensitive volume ensures large signals with no interconnections, low collection electrode capacitance and a low-power, potentially radiation tolerant device [11],[20]. The R&D has now successfully solved the transistor back-gating problem. Prototype thin SOI pixels of 14μm pitch have been successfully operated overdepleted with a 200 GeV π beam at CERN giving a spatial point resolution of 1-2μm and particle detection efficiency above 98% [21].

Based on a more specialized technology, DEPFET pixels have been developed since the late 80's [22] and integrate a p-MOS transistor in each pixel on the fully depleted bulk. Electrons, collected in the internal gate, modulate the transistor current. Their low input capacitance ensures low noise operation and makes them suitable for a broad range of applications from collider detectors [23] to X-ray astronomy [24]. Thick DEPFET devices of pixel pitch of 20-30µm have demonstrated a spatial point resolution of ~1µm [25].

The design of the electronics for monolithic pixels has to be adapted to the limited amount of collected charge and the need to minimize power. In principle, for monolithic pixel sensors collecting charge from the thin epitaxial layer, the Q/C ratio and power consumption might not be necessarily better than that of hybrid pixel detectors. A power consumption and a Q/C improvement can be achieved by minimizing the pixel capacitance C, through a decrease of the size of the collection electrode, and by increasing the charge collection and depletion depth with a moderate- to high-resistivity substrate. Analog and digital power consumption typically differ by less than an order of magnitude in pixel detectors for HEP applications. Analog power reduction by an improved Q/C or by power pulsing has therefore to be matched by a similar reduction in digital power. Power pulsing schemes also for the digital readout are considered, in particular for applications at the ILC due to the low duty cycle. Architectures limiting the off-chip data rate and minimizing the number of operations are also under study. As an example, in the LePix project pursued by an European collaboration centered at CERN [26], the in-pixel circuitry is reduced to minimize the pixel capacitance and data from different pixels are combined in different projections and immediately transmitted to the readout electronics on the periphery of the chip. The projections are chosen carefully to minimize ambiguities. Simulations indicate that occupancies of ~ 50~hits/cm$^2$ can be handled and only 4N signals need to be treated at the periphery for an NxN pixel matrix with four projections. An important component of the power consumption is the off-detector data transmission, and any progress to optimize power/data rate in this area will be beneficial to new applications. Time stamping may be key for reducing accelerator-induced backgrounds and O(1 ns) resolution has been included in studies for multi-TeV collisions at CLIC and a Muon Collider. More relaxed performances of O(10-100 ns) might be applicable for monolithic pixels at LHC and the ILC. The Chronopixel architecture aims at implementing time stamping of ~300ns and data sparsification in the pixel cell of a CMOS monolithic sensor using only NMOS transistors, thus avoiding to depend on special processes [27]. The TimePix architecture, already successfully implemented in hybrid sensors and under study for the LHCb VELO upgrade, may represent an advantageous starting point for the development of a multi-tier 3D sensor attaining ns resolution.

3D integration technologies promise to provide a very elegant way for implementing advanced readout architectures of a pixel matrix [28]. It offers a solution to the conflict between high pixel granularity and fast readout requirements by integrating heterogeneous technologies with individual tiers, one dedicated to charge collection and others providing signal processing and data reduction with dedicated architectures. Digital electronics can be removed from the silicon layer where the sensing electrodes and the analog front-end circuits are located, eliminating interference due to substrate coupling. In a purely digital layer, the designer is allowed to integrate various advanced readout concepts, such as time-ordered readout of hit pixels, in a "data push" or triggered way, enabling a pixel matrix to deal with a hit rate of 100 MHz/cm$^2$ with a readout efficiency close to 100% [29]. There are several approaches to 3D integration differing in terms of the minimum allowed pitch of bonding pads between different layers and of vertical TSVs across the silicon substrate. The HEP community focused a large effort on the design of 3D integrated circuits with the "via first" technology, where the TSVs are etched in the silicon wafers in the early stages of CMOS fabrication [30], provided by Tezzaron/GlobalFoundries[6] with significant leadership provided by Fermilab. With the first functional device now at hand, after several production cycles, this is becoming a very promising approach. Even with less aggressive 3D technologies, such as the so-called "via last" processes where TSVs are fabricated on fully processed CMOS wafers, a significant advantage can be gained in most cases, as only one or two connections are needed between the analog and digital blocks of a single pixel cell, and the digital layer can use low-density peripheral TSVs to reach backside bonding pads for external connection. Devices based on the "via last" technique are being studied within the AIDA EU project [31].

---

[6] Tezzaron Semiconductor, Naperville, IL, USA; GlobalFoundries Inc. Milpitas, CA, USA

Sensors of the thickness routinely achievable with monolithic technologies and back-thinning (~50μm) are compatible with obtaining a material budget of ~0.1% $X_0$/layer, needed to meet the value of b ~ 10μm/GeV for the multiple scattering term in the impact parameter resolution, identified as a requirement at a linear collider and also at second-generation b-factories. For sensors of this thickness, the ladder support is of major importance for counteracting the chip warping and insuring the module stability and planarity. In the design of thin ladders for applications at B-factories, STAR and the linear collider with thin sensors, as well as those for an LHC upgrade, the Si ceases to be the largest single factor in the overall layer material budget which becomes the cable routing signals, power and clocks. Therefore, there seems to be limited interest in pushing the sensor thickness below ~50μm in HEP applications. In order to progress towards modules with overall smaller total material budget, sensor stitching, with clocks and signals routed on metal lines in the chip, appears a promising path. The development of very light double-sided ladders, which are advantageous mechanically and offer hit correlation to reject low energy particles from beam background, appears promising. The PLUME collaboration is addressing some of the challenges related to ladders based on thin monolithic pixel sensors.

# Hybrid Pixel Detector Challenges for next Generation Frontier Experiments at the LHC[7]

Hybrid pixel technology is used at the LHC in several experiments (ATLAS, ALICE, CMS) and it is the baseline for small radius tracking at planned upgrades. Such technology provides the highest radiation tolerance and rate capability available today. The rate capability is a property mainly of the integrated circuit, as are tolerance to ionizing radiation dose and single event upsets. Tolerance to bulk damage radiation dose is a property of the sensor.

The most advanced hybrid pixel system now being built is the ATLAS IBL upgrade that will be installed in the experiment in 2014. The FE-I4 chip meets the specification of ~300 MHz/cm$^2$ data rates and 300 Mrad of ionizing radiation dose while both planar and 3D sensors has been tested and proven reliable up to the required integrated fluence of 5x10$^{15}$ neq./cm$^2$. Moreover the ionizing radiation tolerance of FE-I4 (and therefore the 130nm CMOS process used) has been tested to double the required value.

For a High Luminosity LHC the requirements on rate and radiation essentially triple. For the readout chip the driving challenge is meeting the increased rate (order 1GHz/cm$^2$). This implies a need for smaller and faster pixels and at the same time more in-pixel memory and functionality. A readout chip with a higher density process than 130nm CMOS is therefore needed. Both 3D chip fabrication and smaller feature size CMOS processes have been investigated. The currently favored solution is 65nm CMOS. The main challenges of producing a new chip in this technology are the effort needed to carry out the design and to qualify the new technology to the 1Grad radiation dose level. We expect 65nm CMOS to be rad hard up to such doses, but the qualification work is nevertheless large, because one must, among other things, validate the full digital library in order to enable digital logic synthesis.

A large format pixel IC in 65nm will contain ~500 million transistors, which is more than any single core computer processor ever had (only multi-core processors broke the 1 billion transistor barrier). An IC of this magnitude would need a very large design effort even without considering the extreme radiation tolerance requirement. Such an ambitious goal can only be achieved within our community by pooling resources across experiments and across countries into one focused collaboration. Prototyping in 65nm will also be costly and avoiding effort duplication is therefore important. For these reasons a new CERN RD collaboration is being formed. It will focus on the development of a next generation of pixel readout chips in 65nm. This collaboration will bring together the pixel chip design communities of ATLAS and CMS. This effort will benefit other users of 65nm technology as well. In particular design of pixel chips meeting requirements for ILC or CLIC. A separate white paper on IC design is being prepared, covering all HEP activities.

Purely from a radiation tolerance standpoint, pixel sensor R&D is very active and both 3D and planar pixel technologies are being prototyped to meet the 1.5x10$^{16}$ neq/cm$^2$ challenge. The core of this work is done within the RD50 collaboration. Separate white papers are addressing this. This also falls in the category of challenging, but already being addressed.

While rate and radiation are paramount, they are not the only challenges for the next generation of pixel detectors. There are new ones for the next generation of detectors that were not present in the devices that are presently installed along the LHC ring.

---

[7] Corresponding author(s) for this section: G. Bolla[3], M. Garcia-Sciveres[6]

- An emerging physics requirement is the need for much better 2-track separation in order to maintain tracking efficiency and b-tagging performance for highly boosted jets.
- Reducing mass becomes more important to improve performance in high pileup.
- More efficient power distribution (via on-detector voltage conversion) becomes inescapable to build larger, lower mass detectors.
- Much higher detector output bandwidth will be needed to accommodate higher readout rates (higher rates for the lowest level trigger).

Finally, building larger detectors requires lowering cost and scaling up fabrication capacity. An interesting cost challenge is in fact to realize the potential of flip-chip assembly to be the lowest cost method for fabricating hybrid detectors, as explained later.

Improving 2-track separation requires not only smaller pixels. Already going to higher rate capability required reducing the pixel size, but for separating tracks there is a lower limit to the useful pixel size. This is given by the extent of the charge deposit in the sensor, which is in turn given by the sensor thickness. The 65nm pixel chip design is targeting pixel dimensions of order 25um x 150um. For such pixels to be useful the sensor thickness should not exceed 100μm. While thin sensors well below 100μm have already been fabricated there is still a need for Sensor R&D in this direction in order to establish cost-effective means of mass production for such reduced thickness. Moreover more physics understanding is needed to determine whether this development will fulfill the requirements. Continuing to scale hybrid technology to thinner sensors and smaller pixels is problematic. There is a fundamental limitation. The interconnect capacitance in a bump bonded system sets a minimum value for the noise, hence the sensor thickness cannot be reduced further than S/N allows. To go thinner the only solution is a monolithic system - either a CMOS sensor or 3D integrated chip plus sensor (no bumps).

R&D into radiation hard CMOS pixels is an actively growing area and could become viable on a timescale needed for the HL-LHC. Two approaches are being pursued: the fully monolithic approach of a CMOS readout chip that is also its own sensor, and the hybrid approach where a rad-hard CMOS sensor replaces the transitional diode based planar of 3D sensor. This latter approach takes advantage of all existing infrastructure, but implements a device with an active layer of just 10 to 20μm, solving the S/N problem with internal amplification on the CMOS sensor. Interconnection to the readout chip can also be bump-less using capacitive coupling, which simplifies assembly.

Bump bonding interconnection, which may seemingly also need to scale in order to accommodate smaller pixels, is in fact not a challenge. This is because the technology used for present LHC detectors already has an areal density of 40,000 bumps/cm$^2$ (which correspond to a 50x50 μm squared pixel), which was underutilized. For comparison, the target pixel density of the above mentioned 65nm chip is 27,000 bumps/cm$^2$.

The mass reduction challenge has driven work on mechanics and cooling. $CO_2$ evaporative cooling has been demonstrated and is the baseline for HL-LHC. R&D into other methods continues in order to see if one can still do better. Recent advances in carbon composites include development of thermally conductive carbon foam.

Power distribution work focuses on the two options of serial connection or DC-DC conversion. For the case of hybrid pixel detectors, DC-DC conversion is only low enough mass if integrated inside the readout chip, or if the readout chip power consumption can be reduced by an order of magnitude. In-chip DC-DC conversion is challenging for conversion ratios greater than 2 or 3, due to voltage limitations of the CMOS processes used. However a factor of 2 or 3 is not enough for high efficiency power delivery.

Ideally one would like to read out a pixel detector at the full LHC interaction rate of 40MHz, but this is not considered feasible. One could argue that developing a way to read out a detector with 1GHz/cm$^2$ rate at 40MHz is therefore a challenge that is not being addressed, and would likely be transformative if solved. Readout at 0.5-1 MHz is currently being considered. This would require output data links of 1 or 2 Gb/s per chip for an inner layer at the HL-LHC. This is a practical value for a single serial output in 65nm technology. The main challenge at such bandwidth is from the cables, not the chip design. Optical transmission directly from the chip could also be a solution, but note that the trend in new pixel detectors is to increase reliability by moving optical elements as far away as possible from the interaction point.

In the area of cost, the relevant question to pose is: why aren't hybrid pixel detectors cheaper than strip detectors? The whole point of flip chip bump bonding in industry was to reduce assembly cost relative to wire bonding. The mass production cost of readout integrated circuits per unit area is low. A hybrid pixel sensor provides 2 coordinates with a single sensor, while two strip sensors are needed to do the same job. Sensor costs should be half, and flip chip bonding should be cheaper than wire bonding. It could be that as industrial competition increases, the cost of hybrid pixels will in fact naturally come down below that of strips for the same detector area. It is a challenge to understand if this will happen and prepare to take advantage of it. There are currently several manufacturers in Europe and Japan scaling up bump bonding capabilities. A major factor in the manufacturing cost is the size of the readout chip and the number of chips per sensor, as the flip chip costs scale with the number of chips, not with physical area. The advent of reticle size chips has already greatly lowered the hybrid pixel cost relative those of the first generation of LHC experiments.

# 3D Architecture Pixel Sensors[8]

*Future Requirements and Challenges*

Planned upgrades to the LHC will result in increasingly harsh radiation environments close to the interaction point. A lepton collider would have similar challenges. Existing semiconductor sensor technologies cannot operate effectively over the duration of an experiment under these radiation conditions. One path to meet this challenge is through the use of 3D or non-planar geometry sensors. Development of such sensors has been on-going since the 1990s and is still an active area of study with a growing impact beyond the Energy Frontier. The selection of 3D-geometry sensors for the ATLAS insertable B-layer of pixels during the 2013-2014 shutdown is an indication of the promise of such technologies.  Such sensors can also offer faster charge collection, when for example structured as parallel trenches, than traditional planar sensors. Fast, charge collection maybe also be useful for beam monitors or in time-of-flight systems.

The higher density of tracks resulting from the large multiplicity of interactions per beam crossing will necessitate smaller pixels in order to maintain an acceptable occupancy. To achieve a finer resolution in the Z direction means thinner sensors to prevent low-angle tracks from hitting many pixels. Thinner sensors will also reduce multiple scattering and photon conversions. Since leakage current is proportional to the thickness of the sensor, power and heat generation will be accordingly less as well. The table below summarizes the attributes that will be needed for a vertex sensor a decade from now.

|  | Radiation Tolerance | R-φ Spatial Resolution | Z Spatial Resolution | Radiation Length | Time Resolution | Active Edges | Integrated Cooling |
|---|---|---|---|---|---|---|---|
| Vertex sensors | $1 \times 10^{17} n/cm^2$ | 25 μm | 25 μm | < 0.2 % | < 20 ns | <10 μm | Yes |

*Most promising R&D paths*

Many attributes of 3D sensors must be improved to optimize the system-level performance of future vertexers. Active-edge or thin-edge designs help to minimize material budget and allow for full hermetic coverage. Devices with insensitive borders under 10 microns have been fabricated and confirmed in test beams. This can have a dramatic effect on tracking efficiency. Incorporation of such techniques into production devices is critical.

A significant improvement in radiation tolerance may be achievable by incorporating limited internal gain into the sensor via avalanche multiplication. This would be accomplished by performing extensive calculations to explore the device geometry, biasing, and diffusion parameter space. Prototype runs to confirm the simulation results with follow-on irradiations and beam tests to qualify the concept and final sensors would be needed.

Radiation tolerance could also be improved by reducing the distance between n and p type electrodes. This involves using the most advanced plasma-etching machines and carefully developing the deep-etch recipes to maximize the achievable aspect ratio. Studying techniques to make the electrodes sensitive to ionizing tracks would help both the track efficiency and allow closer electrode spacing.  It should be noted that the 3D architecture has been extended to diamond sensors and that this is an active area of research.

---

[8] Corresponding author(s) for this section:  C. Da Via[18], C. Kenney[12]

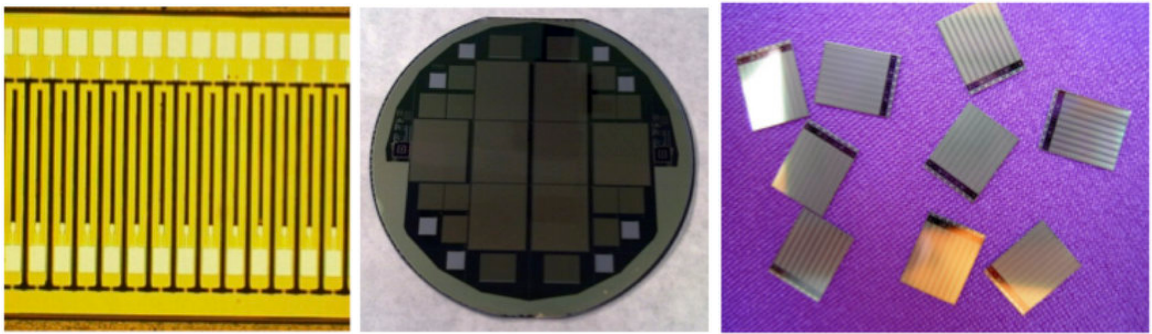

**Figure 1:** Interleaved, fine-spaced trenches for fast response made at Stanford (left), production wafer with ATLAS IBL sensors made at CNM (center), ATLAS FE-I3 3D sensors with active edges (right).

An important feature of 3D sensors are their moderate power dissipation after irradiation, due to the reduced bias voltage needed to get the electric field across the full substrate thickness. This can prove crucial in the design of a low mass system where services, including cooling, might introduce a considerable amount of mass. Micromachining, the same technology used for etching the 3D sensors electrodes, can be used to etch micro-channels, or micro-grooves, in the post process thinned electronics chip substrate, see Figure 1. Images of prototype cooling channels integrated into silicon chips are shown in Figure 2.

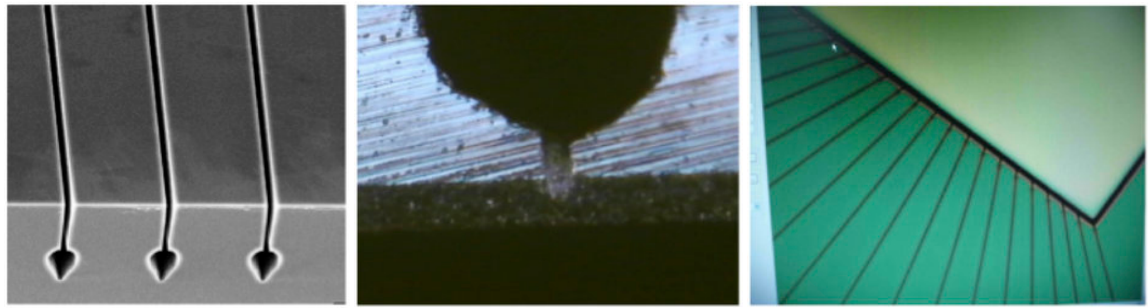

**Figure 2:** Micro-channels etched at FBK for the SUPERB project in collaboration with F. Bosi and co-workers in silicon. The micro-channel is ~50 microns in its larger section. (left) close up showing a glass film sealing the micro-channel aperture. (center) micro-cooling channels emptying into a input manifold made at Stanford (right).

### *Scalability and Costs*

In preparation for construction of the ATLAS insertable B-layer, a consortium of foundries (e.g. CNM, FBK, SINTEF, and Stanford) was organized and has proved very successful in scaling up production of 3D sensors. As the community gains more experience the yields have gone up dramatically with a concordant decrease in cost per unit of sensor area. The micro-machining techniques employed in their fabrication have become ubiquitous in industry and so the opportunity to further partner with commercial vendors is growing.

# Diamond Sensors[9]

Luminosity monitors, beam monitors and tracking detectors in future collider experiments, including LHC upgrades must operate in radiation environments several orders of magnitude harsher than those of any current detector. As the environment becomes harsher detectors not segmented, either spatially or in time, become challenged to separate signal from background. Present tracking detectors close to the interaction region are based on highly segmented silicon sensor technology. Chemical Vapour Deposition (CVD) diamond has a number of properties that make it an attractive alternative for high-energy physics detector applications. Its large band-gap (5.5 eV) and large displacement energy (42 eV/atom) make it a material that is inherently radiation tolerant with very low leakage currents and can be operated near room temperature simplifying the mechanical support and cooling necessary. The drawback of using diamond is that it only provides 60% of the number of charge carriers, per unit radiation length of thickness, relative to silicon. However the steady luminosity increase at colliders, in particular at the LHC, justify considering diamond as a replacement for Si in the innermost layer(s) of the vertex trackers. CVD diamond is being investigated by a number of groups for use very close to LHC interaction regions, where the most extreme radiation conditions are found. Here we describe the requirements for inner tracking sensors at future colliders, highlight the R&D on diamond sensors underway and conclude with a discussion of scalability and cost.

In order to continue the physics studies at the LHC it is imperative that precision tracking continue to be present at the smallest possible radii. In practice, this means the sensors must continue to operate efficiently after fluences of $10^{16}$ charged particles per cm$^2$. At these inner radii minimizing the amount of material in the tracker is paramount as excess scattering and pair conversion significantly compromise pattern recognition and further exacerbates occupancies. Diamond sensors do not require extensive cooling and, being low Z, present less material per unit thickness than silicon or other solid-state tracking media. Of course, to be viable it must be possible to pixelate the diamond sensors, read them out with integrated VLSI electronics and produce reasonably large arrays affordably.

Table 1 shows a list of sensor requirements that would make diamond a viable alternative for future energy frontier trackers. In what follows we describe the R&D aimed at demonstrating that diamond is a viable sensor material for these applications. Although diamond sensors have recently been targeted at collider detector applications, their first use in HEP was in B-factory background monitoring. They continue to be good candidates for beam diagnostics at intensity frontier machines.

Significant progress in diamond sensor R&D has been made over the last five years as the LHC experiments have come on line. All four LHC experiments include diamond beam monitors to protect them from un-stable beam conditions. The most sophisticated of these, in the ATLAS experiment, are also used as the default luminosity counters providing collision and background counting rates that have been stable to better than 1% over the first three years of LHC operation. Reading out the diamond with sub-ns shaping-time electronics has been crucial to achieving this stability in ATLAS as it allows a clear distinction between signals in-time with bunch crossings and background particles that arrive out-of-time. The RD42 collaboration has performed extensive irradiations of CVD diamond material with protons ranging from 26 MeV to 24 GeV, pions at 300 MeV and reactor neutrons. Figure 3 shows the change of carrier mean free path (signal produced when MIPs traverse the sensor) in state-of-the-art CVD diamond material (typical mean free paths of 300 microns in pCVD material) showing superior radiation tolerance to fluences beyond $10^{16}$.

---

[9] Corresponding author(s) for this section: H. Kagan[7], W. Trischuk[8]

| Requirement | Specification |
|---|---|
| Sensor size | 5 x 5 cm$^2$ |
| Sensor Thickness | 400 microns |
| Minimum charge mean free path | 250 microns |
| Minimum average charge before irradiation | 9000 electrons |
| Minimum mean free path/ charge after 10$^{16}$ cm$^{-2}$ | 150 microns/ 5400 electrons |
| Minimum signal/threshold after 10$^{16}$ cm$^{-2}$ | 3 |
| Single hit efficiency | >99.9% |
| Spatial resolution | <14 μm |
| Maximum operating voltage | 1000 V |
| Maximum total leakage current (@1000 V) | 100 nA |

**Table 1:** Diamond sensor specifications

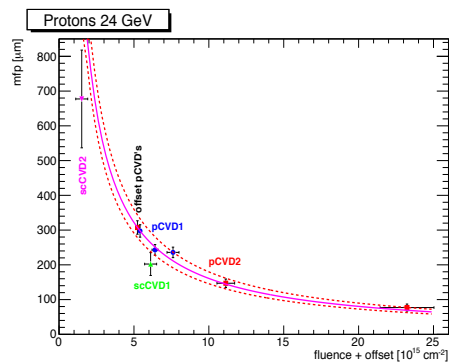

**Figure 3:** Mean free path of charge carriers in CVD diamond sensors vs. fluence of 24 GeV protons. All types follow a common degradation due to the appearance of additional traps. pCVD material is offset to reflect the level of traps in its as-grown state.

Over the last five years there has also been good progress on scaling up the growth of CVD diamond material. Several manufacturers have demonstrated the capability of growing wafers 10cm in diameter (or larger) with charge collection distances that are suitable for single MIP tracker applications. As a result prices have started to come down although they still remain higher than for silicon sensors – although silicon sensors that might survive the dose necessary to survive in future LHC pixel detectors are more expensive due to the elaborate processing they require. In addition to learning how to grow the materials manufacturers have learned that surface preparation is important to the overall quality of the sensors and the long-term survival of electrical contacts to the resulting sensors. The CMS experiment is preparing a set of forward pixel telescopes, based on the largest single crystal CVD diamond sensors currently available (0.5 x 0.5 cm$^2$) for installation during the current LHC machine stop. In turn ATLAS is preparing a pixel Diamond Beam Monitor using larger poly-crystalline CVD diamond sensors (1.8 x 2.1 cm$^2$) on a similar timescale. Both of these projects have demonstrated routine large-scale bump-bonding of VLSI pixel readout chips to diamond sensors and aim to demonstrate routine tracking in the most demanding environment yet explored, less than 5 cm from the LHC beamline and 1m from their respective, LHC interaction points. A number of studies have been performed to characterize the tracking performance with diamond sensors. Charge sharing has been demonstrated with the diamond strip trackers giving position resolutions of better than 9μm [33] and similar resolution has been seen in 50μm wide pixels in small single crystal pixel devices [34]. Resolution has been shown to improve in polycrystalline CVD sensors after irradiation as the introduction of defects improves the material uniformity [35].

The development of diamond sensors for HEP applications has piggy backed on industrial production capabilities for other applications. CVD substrates have applications in thermal management, X-ray windows and, potentially, dosimetry (with Carbon being a better match for human tissue than other solid-state sensor materials). Over the past ten years the RD42 collaboration has worked with a dozen different manufacturers. At the present time we have two viable suppliers (Element6 and II-VI Ltd.). Both have the capability to produce tracker quality substrates and have demonstrated the capacity to produce several hundred cm$^2$ per year of sensor material (see Figure 4). Figure 5 shows the size of the various diamond systems that have been installed over the last 10 years. The current capacity, and the potential manufacturers have to increase that capacity, should be sufficient to build two or three tracking layers at radii up to 10cm in both the multi-purpose LHC experiments prior to the ultimate luminosity running in the next decade.

Diamond sensors have proven highly reliable replacements for silicon in every HEP application they've been used in over the past 10 years. In each case they've allowed measurements that were un-expected when they were first proposed. We expect that they will continue to revolutionize the way we do tracking in high occupancy, high radiation environments at the high luminosity LHC and beyond.

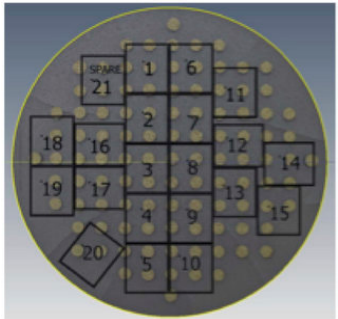

**Figure 4:** A poly-CVD diamond wafer from II-VI Inc. for the ATLAS Diamond Beam Monitor, showing 21 sensor substrates prior to dicing.

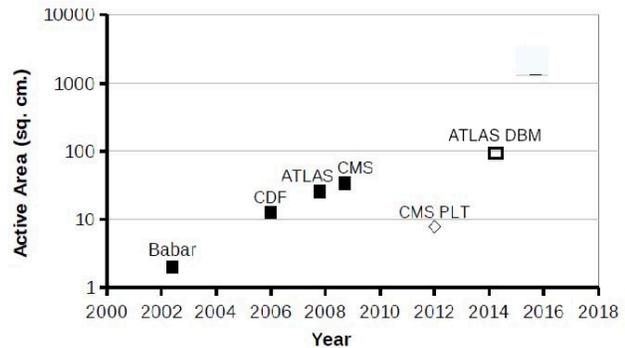

**Figure 5:** Active area of various diamond detector systems that have come in to operation over the last ten years. The open symbols show tracking devices that are currently under construction.

# Ultra-fast Silicon Detectors (4D-UFSD)[10]

## 1. *Principle of 4D-UFSD*

The ultimate goal of 4D-UFSD is to establish a new paradigm for space-time particle tracking by developing a single device that matches the performances of the most accurate position and time detectors. The goal is to obtain a position resolution of ~10 μm while measuring the arrival time of the particle with an uncertainty of ~10 ps. In the past, precise tracking devices have determined time quite poorly while good timing devices are too large for accurate position measurement.

The goal is to develop the 4D capability by augmenting the superb position resolution of a segmented silicon sensor with high-resolution timing information.

The 4D-UFSD principle is based on

a) Use of thinned silicon pixel/strip sensors to reduce the charge collection time.
b) Increase the charge presented to the read-out electronics by employing charge multiplication within the sensor bulk.
c) Develop a read-out system that will match the silicon sensor rate capability, segmentation and time resolution.

## 2. *Gain in Silicon Sensors*

Given that the drift velocity in silicon saturates at about $10^7$ cm/sec, the collection time of electrons inside a silicon layer of ~300 μm is ~3 ns. Much faster silicon sensors need therefore to be very thin which means that the charge collected from a thin active layer might not be sufficient to achieve good time resolution. We propose to exploit charge multiplication to increase the charge yield of very thin silicon sensors so that they can generate ultra-fast timing signals [36]. The past observation by several RD50 groups [37] of moderate gain in irradiated silicon sensors, and the new measurements on un-irradiated pad detectors with optimized implants showing gain of about 10 [38],[39] creates the opportunity for producing thin, very fast silicon sensors, which can work at extremely high-rates without dead-time issues.

It has been demonstrated that in silicon sensors the charge multiplication factor $\alpha$, which is responsible for the charge gain, has an exponential dependence on the inverse of the electric field [40]. At the breakdown field, $E_{max}$= 270 kV/cm, $\alpha$~0.7/μm for electrons and $\alpha$~0.1/μm for holes, which limits the maximum achievable gain g. $N_0$ electrons, drifting a distance *d*, with a charge multiplication factor $\alpha$ become $N_{Tot}$ electrons:

$$N_{tot} = N_0 e^{\alpha * d} = g N_0$$

For example, electrons drifting 5 μm in the maximum field generate a gain g = $N_{Tot}/N_0$= 33. Changing the field by 5% changes the gain by more than a factor 2.

## Sensor Thickness: Capacitances, Collected Charge, Collection Time, Gain

The thickness of the active area determines several key parameters of the sensor. Considering two possible geometries, (i) a 50 μm x 50 μm pixel and (ii) a one mm-long strip with 50 μm pitch, Table 2 shows the backplane capacitance, the number of electrons that form the signal based on [41], the collection time, and the gain required to reach an acceptable signal level, 2000 electrons for the pixel, and 12000 electrons for the strip sensor [42].

---

[10] Corresponding author(s) for this section: H. Sadrozinski[2]

Taking values of the inter-strip capacitance for existing pixel sensors (200 fF) and n-on-p strip sensors (1 pF/cm) as a guideline, a 4D-UFSD thickness where the backplane capacitance would not dominate are 2 μm for pixels and 5 μm for strips.

From the values shown in Table 2, pixel sensors offer very attractive combinations of moderate gain, small capacitance, and short collection time. Due to the high value of backplane capacitance, strip sensors cannot be made as fast as pixel sensors; however, a "mini-strip" of 1-2 mm long offers a quite fast collection time, 50-100 ps, with a moderate value of capacitance (~1 pF).

| Thickness | Backplane Capacitance | | Signal | Coll. Time | Gain Required | |
|---|---|---|---|---|---|---|
| [μm] | Pixels [fF] | Strips [pF/mm] | [# of e$^-$] | [ps] | for 2000 e | for 1200 e |
| 2 | 125 | 2.5 | 80 | 25 | 25 | 149 |
| 5 | 50 | 1.0 | 235 | 63 | 8.5 | 51 |
| 10 | 25 | 0.50 | 523 | 125 | 3.8 | 23 |
| 20 | 13 | 0.25 | 1149 | 250 | 1.7 | 10.4 |
| 100 | 3 | 0.05 | 6954 | 1250 | 0.29 | 1.7 |
| 300 | 1 | 0.02 | 23334 | 3750 | 0.09 | 0.5 |

**Table 2:** Silicon sensor characteristics for various thicknesses of the active area.

## Sensor Processing Options

An important research activity is the optimization of the sensors for stable processing and operations. This includes efforts to both better quantify what is physically happening in the sensor and to tailor the field to optimize the performance for timing purposes and for minimum material. The gain phenomenon in segmented silicon sensors is in need of deeper understanding, and then to translate this understanding into processing.

 a. **Wafer type p-on-n vs. n-on-p**

Several arguments favor n-on-p sensors (which collect electrons) over p-on-n (which collect holes): the collection of electrons is faster, electrons have a factor of 10 larger multiplication gain [40] and there have been many studies on radiation-hardness of p-type sensors for the LHC upgrade [42][43][44].

 b. **Epitaxial and Thinned Float Zone**

Thin epitaxial sensors are easily produced since they consist of a low-resistivity electrode (n$^{++}$) implanted in a high-resistivity p$^-$ epitaxial layer of silicon, deposited on a thick low-resistivity p$^{++}$ substrate. Given the low costs and quick manufacturing turn-around time, thin epitaxial sensors are an ideal low-cost vehicle for prototyping, e.g. for optimizing the field configuration and other key parameters. Ultimately one wants to eliminate the thick "handle wafer" and migrate to thinned FZ sensors. This process uses a FZ wafer of the usual thickness of several hundreds of micro-meters and removes the excessive material at the backside by etching. Several groups have produced back-etched thin sensor, down to 15 μm thickness, using a selective etch to create ribs for mechanical strength [45][46].

### c. Planar and 3D

Planar sensors collect charges in implants on the surface, while 3D sensors [47] collect charges in columns implanted in the sensor bulk. For 3D sensors, there are good theoretical arguments [48] and supporting their use as fast sensors with gain. The main problem might be that it would be difficult to achieve the tuning of the field around the narrow columns with the precision described in the next section.

### d. Tuning of the Electrical Field

The field profile in the sensors is determined by the doping profile and the bias voltage. Fine tuning of the doping profile is required to simultaneously satisfy the different requirements.

#### i. Minimum Field

One of the requirements is that in order for the electrons to be moving at the highest possible velocity, the field needs to exceed 25 kV/cm, the field required for saturating the drift velocity [49].

#### ii. Depth Doping Profile

In February 2013, RD50 groups [38],[39] have reported gain in un-irradiated pad sensors using specialized deep n++ implants with underlying p+ diffusion [50] (Figure 6). Adopting the same principle to thin sensors, one can establish a high-field region close to the collecting n++ implant providing the charge multiplication, with a somewhat lower field region in the back of the sensors providing the charge to be amplified. Figure 7 shows the principle of such a field configuration. An important property of this figuration is that the "amplification field" strength is essentially independent of the bias voltage, thus providing for stable operation and low currents. Charges generated in the bulk and drifting to the n++ implant have to pass through this high filed area and will be amplified. Tuning the doping density within this amplification field will be a major R&D task.

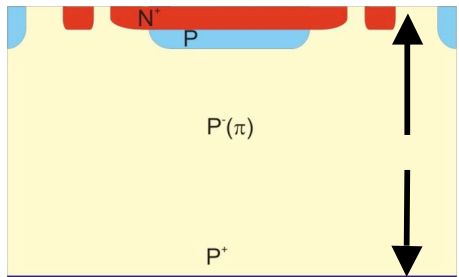

**Figure 6:** Simplified electric field distribution in a 20 µm thick p-type sensor with a 100 Ohm-cm bulk and 4 Ohm-cm p+ diffusion. The drift field is above 25 kV/cm with only small over-voltage above 125V , and the "amplification field" is constant for large over-voltage.

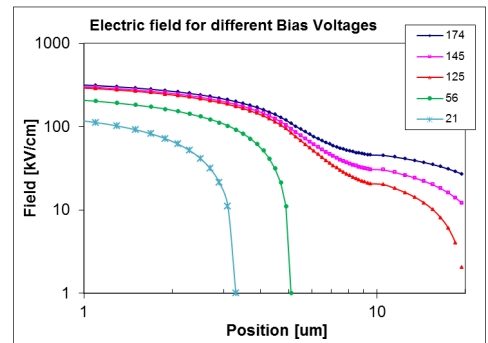

**Figure 7:** Schematic of the deep p+ diffusion below the n++ implants for charge multiplication silicon sensors [50]

#### iii. Lateral Doping Profile

Moving from pads to segmented sensors and pixels raises the problem of lateral field uniformity across the n++ implants. Without careful design of the doping profile, charge multiplication happens only in small parts of the sensor, for example near the implant edges where the field lines become very dense. Thus only the charges arriving at the implant edges would be amplified at moderate bias voltages, and when the bias is raised to achieve charge multiplication at the center of the implants, the strip edges would experience electrical breakdown. It is therefore necessary to change the geometry of the field lines inside a sensor in order to prevent this problem, while achieving a large enough volume with high field. By tailoring the depth and/or doping concentration of the additional $p^+$ diffusion layer, it is possible to adjust the maximum electrical field across the strip width, which can lead to approximately equal multiplication mechanism across the electrode.

# Next Generation Strip Detectors for the Energy Frontier[11]

The introduction of Silicon Microstrip detectors in particle physics has been critical for many major discoveries from the top quark to the Higgs. Future experiments operating at high-energy colliders will continue to rely on this technology since the robust tracking and detailed vertex reconstruction that can be achieved with solid state detectors are of central importance to the physics that can be accessed at future hadron colliders including the HL-LHC and the HE-LHC. Solid state detectors can maintain low occupancy, and provide precise and efficient determination of charged-particle momenta which impacts the measurement of all physics objects from leptons, to charged hadrons, jets, and photon conversions. In order to be sensitive to physics involving the third generation and higher mass objects we must achieve efficient b tagging for high momentum jets.

## *1. Low mass mechanical supports and cooling systems*

The material in the current tracking systems pose one of most severe limitation on the performance of the detector. It is often dominated by electronics and service components and for example in CMS it presents a major problem especially in the transition region between barrel and end-cap. Furthermore at the HL-LHC the number of particles produced is expected to increase substantially, causing higher occupancy, and increased radiation damage. As a result the power consumption of new trackers is expected to be even larger. $CO_2$ cooling systems are expected to enable a reduction of material inside the tracking by providing an effective and low mass alternative to the current cooling infrastructure. R&D for the development of new integrated approaches, including powering, and cooling is therefore critical for the HL-LHC.

ATLAS has developed a strawman design of the strip system for the HL-LHC [51]. The stave (petal) is the basic mechanical element of the barrel (end-cap). The barrel module consists of a low mass central core that provides mechanical rigidity and houses the common electrical, optical and cooling services. In order to reduce the material, modules are constructed by directly gluing kapton flex hybrids to silicon sensors with electronics-grade epoxy. The design uses the large cross-sectional area of the hybrid and sensor to the stave core for cooling. The cooling pipes are embedded in a carbon fiber honeycomb sandwiched between two carbon fiber facings. The carbon fiber has both high tensile modulus and high thermal conductivity. The % radiation length is 1.98 for the barrel (1.60 for the end-caps) in comparison with 2.48% for the barrels (3.28% for the end-caps) for the HL-LHC and the current SCT respectively.

ATLAS is conducting R&D to develop a reliable and low-mass connection method for the several thousand $CO_2$ connections needed [51]. Several options are being investigated, including direct orbital welding of thin-wall Ti tubes; brazing stainless end-pieces to the Ti and welding to standard stainless welding connectors; and developing low mass in-house connectors that can be disconnected and re-used. For in-situ welding, tests are under way to ensure the front-end electronics are not damaged by the process.

CMS is also conducting R&D on the mechanical modules [52],[53]. The concepts under development enable the correlations between top and bottom sensors necessary to provide trigger information. They include strip-strip ($p_T$-2S) and pixel-strip ($p_T$-PS) modules. The current implementation of the $p_T$-PS modules has an average radiation length of 2%. A more elegant but challenging approach to the construction of a pT module is represented by the vertically integrated Pixel-Strip module (or $p_T$-VPS). In this design one 3D chip reads out the pixelated sensor and the short-strip sensor, connected by analogue paths through an interposer, and implements the correlation logic. The use of vertical connectivity removes all constraints between module dimensions and sensor segmentation. On the other hand, the interconnection technologies are challenging: feasibility, reliability and yield for large-surface assemblies need to be verified, especially considering the demanding operating conditions inside CMS. In addition, in this concept an interposer covers the entire surface of the module, defining the spacing between the two sensors, providing the top-to-bottom connectivity and carrying at the same time power lines and readout signals. The realization of a lightweight interposer with the needed thickness (about 1mm) and electrical and mechanical properties is a key issue to realize a module with an acceptable material density.

---

[11] Corresponding author(s) for this section: D. Bortoletto[3]

## *2. Electronic Readout*

The R&D for the readout of the HL-LHC strip tracker is currently focused on the 130 nm CMOS process. The CMS CBC2 ASIC [54] is designed to instrument double layer modules in the outer tracker, consisting of two overlaid silicon sensors with aligned microstrips, and incorporates the logic to identify L1 trigger primitives in the form of "stubs" as described in section 5. The ROC will adopt bump-bonded C4 pads, so that routing and pitch adaptation of the input signals from the sensors can be made in an advanced hybrid technology. The logic analyses on adjacent strips from the same sensor allow rejecting wide clusters of hits associated with low-momentum tracks. The coincidence between inner sensors and outer sensors can be adjusted according to the radii and the position of the strips along the module. Care has been taken to keep the power consumption low. The power consumption of the stub finding logic depends on occupancy, maximum cluster width and coincidence window width; however simulations in pessimistic conditions (4% strip occupancy, maximum cluster and coincidence widths) indicate a value of less than 50 $\mu$W/channel. A total power of less than 300 $\mu$W/channel for 5 pF load has been achieved. ATLAS has prototyped the ABC130 (130nm CMOS ASICs) [55] which is compatible with both serial powering and DC-DC convertors.

A challenge of the electronic chain is the transmission of large volume of data. Both ATLAS and CMS are currently planning to use the GigaBit Transceiver (GBT) radiation-hard bi-directional 4.8 Gb/s optical fiber link [56] to transmit the data between the counting room and the experiments. Currently R&D is needed to provide a version of the low power version of the GBT which will be designed using 65 nm technology and will only use ¼ of the power of the current GBT, i.e. but otherwise will use the same protocol and achieve the same speed.

## *3. Power Distribution*

Future Collider trackers require large granularity and therefore the power delivery problem is especially challenging. Both serial powering and DC-DC powering are being considered by ATLAS while CMS is focusing only on the latter. Serial powering uses shunt regulators on each module to generate the voltages needed by the front-end ASICs. A Serial Power Protection ASIC (SPP) is being designed [57] that in combination with pass transistors built into the ABC130 will act as a shunt regulator. Prototype staves have been built also with commercial shunt regulators.

DC-DC conversion is a realistic option for improving power delivery efficiency. A buck converter is a circuit topology that can transform DC power using only one inductor. The energy is pumped into the magnetic field at the higher voltage and withdrawn at the lower voltage. If the buck converter operates at a high frequency, the required inductance can be quite small, and an air core inductor can be used in high magnetic fields. CMS is already producing buck converter prototypes, named AC PIX V8, based on the most recent ASIC prototype, AMIS4 [58]. A toroidal plastic-core inductor with an inductance of 450 nH is used, and the switching frequency is configured to 1.5 MHz. The fully populated converter boards have a weight of about 2 g. The DC-DC converters are equipped with a shield.

DC-DC regulators are used for a large variety of commercial applications and therefore large R&D investments have been made by the semiconductor industry over the last 50 years. The industry continually strives to produce more efficient and higher frequency power semiconductor devices. Furthermore portable devices such as smart phones must produce several different voltages and therefore they require very small, lightweight DC-DC converters. Currently several companies are developing Gallium Nitride on Silicon which will enable DC to DC converters operating at higher frequencies than Silicon devices. GaN devices are expected to be rad hard and will be used in space and satellite hardware and in military electronics. It is important for future experiments to look if some of the industry advances can be used effectively in future trackers [59].

## *4. Sensor and Radiation Hardness*

One of the key issues for strip sensors deployed in a future tracker is to withstand fluences of the order of $2\times10^{15}$ $n_{eq}$ $cm^{-2}$. While these specifications are not as stringent as for pixel detectors, R&D is still necessary in order to find cost effective radiation hard sensors. Dedicated R&D has been conducted both by ATLAS and CMS. The CERN-RD50 collaboration has also been instrumental in the study of the radiation issues for the HL-LHC.

CMS is conducting a complete measurement campaign to qualify sensor technologies suitable for the HL-LHC [60]. Sensors on 6-inch wafers were produced by Hamamatsu Photonics with different substrates, thickness, implants and geometries. Three different technologies for the substrate production are investigated, with different active thicknesses: Float-zone (120, 200, 300 µm), Magnetic Czochralski (120 µm) and epitaxial (50, 100 µm). For each substrate technology wafers were produced with three doping types (n-type bulk, p-type bulk with p-spray or p-stop isolation). Multi-geometry strip detectors are the largest structures with 3 cm-long strips. Arrays of long pixels (approximately 1 mm long) that can be read out by a bump-bonded chip and used to directly measure the track z coordinate are also been implemented.

ATLAS has focused the R&D on strip sensors that are AC-coupled with n-type implants in a p-type float-zone silicon bulk (n-in-p FZ). These sensors collect electrons and they do not undergo radiation induced type inversion. The standard sensors are about 320 ±15 µm thick, but the sensor thickness could be reduced to 250 µm for a higher cost [51].

The initial measurements that have been conducted indicate that the n-on-p sensors match the requirements of the HL-LHC. There has also been progress in the understanding of the field distribution inside heavily irradiated sensors using Edge-TCT [61] to extract the velocity profiles. This can help developing a precise method for predicting the electrical performances of the devices up to the maximum fluence anticipated for a given experiment and to accurately predict expected performances of heavily irradiated sensors in term of signal, noise and evolution after irradiation.

## *5. Intelligent Trackers*

As the luminosity of accelerators increases it becomes increasing challenging to maintain the present level of physics performance of the detectors and the low trigger thresholds necessary for studying the Higgs boson. Therefore new triggering and data collection approaches are required to improve filtering and data throughput. Several approaches are under considerations.

One option aims at reducing the volume of tracker data to be read out with the definition of a Region-of-Interest (RoI) set by the calorimeter and muon Level-0 triggers. This option can only be implemented in a split Level-0/Level-1 trigger architecture. This is the baseline design for the ATLAS L1 track trigger for the HL-LHC [51]. The RoI-driven L1Track trigger would provide all tracks in a number of relatively small regions of the inner tracking detector. The full data from the tracker cannot be read out at the full beam-crossing rate or even at a Level-0-Accept rate of 500 kHz. The regional readout design of the track trigger addresses this problem by starting from the Level-0 trigger RoIs and then reading only a subset of the tracker data. Conceptually, this is similar to the current design of the ATLAS Level-2 trigger, FTK [62], except that between the Level-0-Accept and the Level-1 decision the data are buffered on detector in the inner detector readout electronics. One of the main advantages of the RoI-driven L1Track design is that it has little impact on the inner detector layout that, apart from the depth of the front-end buffers, can be optimized almost independently from the design of the trigger. The main challenge for the RoI-driven L1Track trigger is to fit within the ATLAS Phase II trigger latency constraints.

A more challenging concept relies on finding track segments that would be used as primitives for a self-seeded track trigger [63],[64],[65]. This requires specially designed and instrumented trigger layer pairs that would be used to identify high transverse momentum (high-$p_T$) candidate track segments with low latency to seed the first level track trigger with candidate collisions for full readout. The rejection of low $P_T$ tracks relies on correlating signals in two closely spaced sensors since the distance between the hits in the x–y plane is correlated with the curvature of the track the magnetic field and therefore $p_T$. A pair of hits that fulfills the selection requirements is called a "stub". This method places constraints on the geometrical layout of the tracker. For a given $p_T$, the distance between the hits forming the stub is larger at larger radii. If the module is placed in end-cap configuration, the same discriminating power is obtained with a larger spacing between the two sensors, compared to a barrel module placed at the same radius. CMS is currently considering a long barrel and a barrel plus end-cap layout. . In order to implement isolation cuts in the Level-1 selection, the Tracker will have to provide information on all tracks above 2 GeV, with a precision in the position of the collision vertex along the beam axis of at least 1 mm.

Two different architectures for the CMS L1 track reconstruction are also considered. One using hits with a small cluster size and pattern recognition with a content addressable memory (Associative Memory, AM) which compare the hits with pre-stored track patterns. Research and development is necessary to develop new generations of CAM chip regardless of whether such a chip will be used for efforts such as an FTK upgrade or for the L1Track pattern recognition [66].

In a second implementation that favors a long barrel geometry, the transverse momentum of charged particles is determined from the correlation between stacked double strip layers. The stacked trackers are composed by one or more double pixel layers, spaced approximately one millimeter apart. A pair of stacks spaced a few centimeters apart is called a "double stack". When hits are registered on both stack members within a stack and the separation in of these hits is consistent with a high-$p_T$ particle, the set of hits become known as a "stub", the basic trigger primitive of a stacked-tracker geometry. Correlated stubs on two or more stacks are called a "tracklet". Pattern-based hit correlation of properly built clusters of hits provide L1 trigger primitives to the hardware trigger. These can be combined together in a projective geometry to perform a rough tracking to be implemented online, returning rough $p_T$, direction, and vertex information for a candidate track. The modules required implementing this approach benefit by using 3D connectivity, which incorporate through-silicon-vias to provide correlation between the tiers to find stubs [53].

In general R&D is needed to improve the link technology and transfer large amount of data in shorter times. Recently the first prototype of a 60 GHz wireless Multi-gigabit data transfer topology is currently under development at University of Heidelberg using IBM 130 nm SiGe HBT BiCMOS technology. This could deliver multi-gigabit per second data rates, and therefore it could be a potential solution to the challenging data transfer rate from highly granular tracking detectors.

## 6. Summary

Even if the requirements on the ATLAS and CMS strip trackers are not as severe as for the pixel detectors, there are many issues that remain very challenging. Furthermore since the trackers for the CMS and ATLAS upgrade will cover large areas, special attention must be places on integration, powering, cooling, and material support. The R&D areas are summarized in Table 3.

| Issues | System | Sensors | FEE | Optical links | Mechanics & Cooling | Powering |
|---|---|---|---|---|---|---|
| Radiations Hardness | | $2E15$ n/cm$^2$ | 130 nm | 65 nm | Certify materials | DC-DC / serial power |
| Reduce % radiation Length | | Low cost thinner sensors 250 $\mu$m | Reduce power/ channel | Low power links | Low mass support material CO2 cooling | Low mass components |
| Occupancy | 1% | Reduce Strip length/ introduce micropixels | | High bandwidth | | |
| Resolution | | Pitch 50 $\mu$m | | | Alignment | |
| Connectivity | | C4 bonding | | | | |
| Tracker information for level 1 | | Edgeless, Slim edge | | Increase bandwidth | TSV 3D electronics | |
| Integration | Facility to build and access | Reduce type of sensors | | | Reduce Module types | |

**Table 3:** Requirement for silicon microstrip detectors for the HL-LHC upgrade

# Next Generation Strip Detectors for the Intensity Frontier[12]

Silicon strip sensors are now the standard building block of charged particle tracking detectors in several different applications including Intensity Frontier experiments. We will summarize the purpose, goals, and challenges of the next generation of such devices being currently designed.

Silicon microstrip detectors for "intensity frontier" applications must meet requirements that are similar in nature, although quantitatively different, than the ones planned for "high energy frontier" applications. Two different implementations are discussed here, one being planned for the LHCb upgrade and the other for Belle II. The main requirements are summarized in Table 4.

| Requirement | LHCb UT [68] | Belle II SVD [69] |
| --- | --- | --- |
| Radiation fluence expected | $2 \times 10^{14}$ $n_{eq}/cm^2$ | |
| technology | Single sided n-in-p or p-in-n | Double-sided |
| pitch | 100-200 mm | |
| Strip length | 5-10 cm | |
| Fan-out | Double metal layer or | Double metal layer |
| Front end electronics Analog section peaking time | 25 ns | 50 ns |
| Front end readout scheme | Zero suppressed digital 3-5 bit ADC | Analog pipeline non-zero suppressed |
| Average radiation length | <1%$X_0$/layer | 0.58%$X_0$/layer |
| Power dissipation | | 0.35W/128 channels |

**Table 4:** Requirements for silicon microstrip detectors for the UT Tracking Module in the LHCb upgrade and for the SVD of Belle II

Although the experimental environments in which these two detector operate are different, similar requirements need to be addressed:

1. The Belle II detector, in order to maximize the spatial information achievable with a single silicon layer, utilizes double sided readout to extract more information from a single Si layer.

2. The LHCb design is including a "shaped inner edge," similar to the VELO design, to achieve closer proximity to the beam pipe [70].

3. Radiation resistance considerations are relevant to both detector systems, although they are much more demanding in a hadron machine. In the case of the LHCb upgrade it is important to withstand highly non-uniform irradiation. In addition, as fan-out of the signal connections may be implemented through a double layer of metallization, possible radiation effects induced by this layer need to be established [71].

---

[12] Corresponding author(s) for this section: M. Artuso[1]

4. A minimization of the overall material budget is critical to optimize tracking efficiency and resolution. This can be achieved only with a design that tightly integrates the electrical, cooling, and mechanical specifications.

5. In order to cope with the higher occupancy (Belle II) and to achieve the necessary data acquisition speed (LHCb), fast shaping times are needed.

6. The requirement of pushing the data out of the detector at 40 MHz implies that the LHCb tracking system needs to implement the zero suppression algorithm in the front end ASIC.

**Major R&D Items:**

1. Front end electronics for fast data acquisition system
2. Low mass system design (synergistic with the chapter on low mass mechanics and cooling)
3. Low mass interconnection scheme preserving fast data rate
4. Qualification of all the components for the necessary level of radiation resistance.

# Low Mass Support and Cooling for Future Collider Tracking Detectors[13]

## *1. Introduction:*

Future high luminosity (HL) and high energy (HE) colliders present a variety of technical challenges for the development of low mass support structures and cooling systems. These challenges derive from the need to operate large area sensor arrays with high channel counts, at low temperature, and in high radiation fields.

Physics reach is being extended in both HE and HL colliders due to increased cross section and interaction rates. These both lead to increased particle multiplicity. As a result the innermost tracking detectors, typically silicon but potentially diamond or other materials, are subject to unprecedented radiation dose exposures. To decrease detector leakage currents (and hence electronic noise), and stabilize depletion voltage changes, these detectors will need to operate at ever decreasing temperatures. Increased multiplicity also results in the need for finer detector segmentation and thus higher readout electronics density. Therefore future detectors will have greater power densities that will require efficient cooling. Table 5 lists typical operating or environmental conditions for high luminosity tracking detectors.

| Operating/Environmental Conditions | |
|---|---|
| Max Operating Temperatures | 40 °C |
| Min Operating Temperatures | -50 °C |
| Ionizing Radiation TID | 50 - 1000 Mrad |
| Non-ionizing radiation Levels | $1\text{-}10 \cdot 10^{15}\ n_{eq}/cm^2$ |
| Magnetic Fields | > 2 T |
| Typical Sensor Area | 100 $m^2$ |
| Typical Power Densities | ~ 100mW/$cm^2$ |
| Total Power | 100 KW |

**Table 5:** The typical environmental and operating conditions for HE and HL collider tracking detectors

The size and scale of the detectors at HL and HE must be large to generate enough hit points, over sufficient lever arm, to permit good track and momentum reconstruction. Typical tracking detector sizes can exceed 100 $m^2$ of silicon. The mechanical support must therefore be similar in scale, and consequently must efficiently utilize the support material to reduce overall detector radiation length. This serves to reduce multiple scattering and conversions. Efficient material use requires structures to perform dual (or triple) roles of providing support, cooling, and/or electronic functions such as shielding and signal/power distribution. Use of low-Z materials that have a high stiffness to mass ratio is required.

Table 6 lists general requirements for support structures for HL and HE tracking detector supports.

---

[13] Corresponding author(s) for this section: W. Cooper[9], C. Haber[6], D. Lynn[10]

| Requirements |
|---|
| Long radiation length materials |
| High stiffness/mass ratio (i.e. stability) |
| Radiation tolerant materials |
| Dual (or triple) purpose elements |
| Large scale manufacturability |
| Minimal thermal and humidity effects |

**Table 6:** General requirements for mechanical supports for HL and HE collider tracking detectors.

## *2. Materials:*

In general, low-Z materials are to be preferred for use in support structures as these will typically have longer radiation lengths. Polymers such as plastics have long radiation lengths as well. A material is usually selected for its stiffness and/or thermal conductivity. All materials need to be qualified as radiation hard to levels somewhat higher than the total radiation dose expected over the lifetime of the detector.

Table 7 lists some materials used in tracking detectors for mechanical support and cooling. Their relative merits in terms of stiffness, thermally conductivity, and radiation length are indicated with "✓"s. Generally metals are to be avoided relative to other materials as their performance, with the exception of beryllium, is lower. (Beryllium suffers from high cost and safety issues that limit its use). Materials derived from carbon have found wide use for their overall favorable properties and versatility. Ceramics may be used in limited quantities when simultaneously electrical isolation and high thermal conductivity is desired. Polymers find application when complex shapes are needed (either via machining or injection molding) and stiffness is not the primary driver.

|  | Structural (stiff) | Thermally Conductive | Low (long) Radiation Length |
|---|---|---|---|
| **Carbon based materials** |  |  |  |
| Carbon Foam | ✓ (as core material) | ✓✓ | ✓✓✓ |
| Carbon Fiber Reinforced Plastic | ✓✓✓ | ✓✓✓ | ✓✓ |
| Carbon Honeycomb | ✓✓ (as core material) | X | ✓✓✓ |
| TGP | X | ✓✓✓ | ✓✓ |
| **Polymers/Plastics** |  |  |  |
| Peek | ✓ | X | ✓✓ |
| Liquid Crystal Polymer | ✓ | X | ✓✓ |
| **Ceramics** |  |  |  |
| Alumina | ✓✓✓ | ✓ | ✓ |
| Beryllia | ✓✓✓ | ✓✓✓ | ✓✓ |
| Aluminum Nitride | ✓✓✓ | ✓✓✓ | ✓ |
| **Metals** |  |  |  |
| Beryllium | ✓✓✓ | ✓✓✓ | ✓✓ |
| Aluminum | ✓✓ | ✓✓✓ | ✓ |
| Titanium | ✓✓ | ✓ | X |
| Stainless Steel | ✓✓ | ✓ | X |

**Table 7:** Typical materials and their relative performance in terms of stiffness, thermal conductivity, and radiation length.

| ✓✓✓ | excellent |
|---|---|
| ✓✓ | good |
| ✓ | decent |
| X | poor |

### *3. Structures*

With the focus mainly, and naturally, on carbon based composite materials, a great variety of structures and arrangements have already been studied, and many more may be envisioned for the future. Figure 1 shows a prototypical support and cooling structure fabricated from predominately carbon based materials.  This particular structure targets a two layer pixel tracking system for a HL collider. In application, a group of these would surround the interaction region. The structure is constructed as an I-beam for rigidity. Note the structure's multi-purpose functions. It supports two radii of (pixel) sensors while providing cooling to each. Carbon fiber re-enforced plastic facesheets provide stiffness and thermal conductivity. Carbon foam provides a thermally conductive path from the metal pipes to the facesheets.

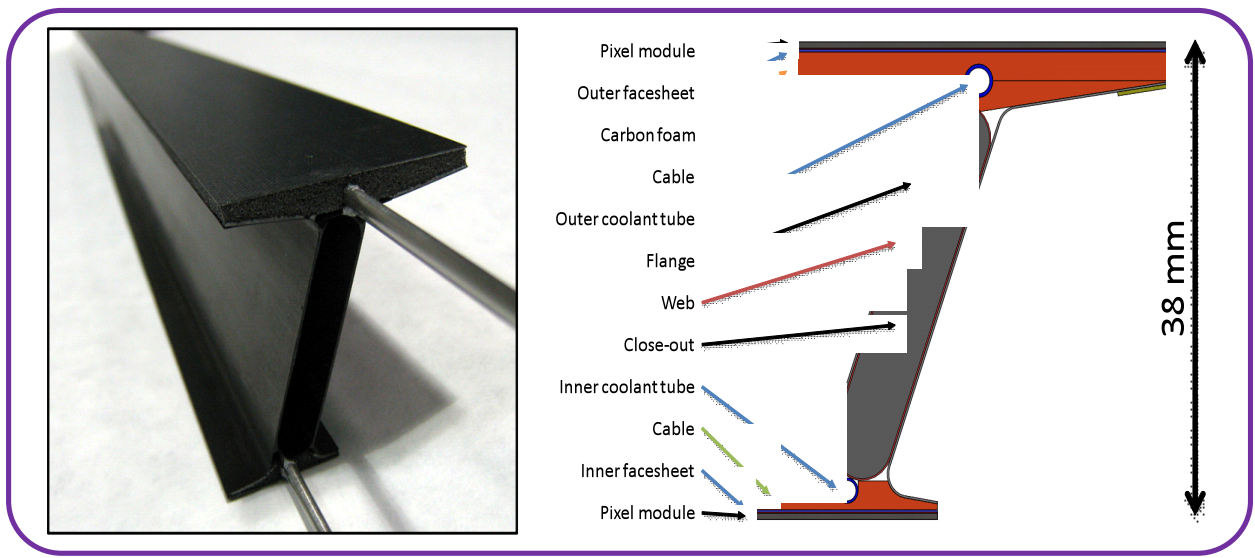

**Figure 8:**  Carbon composite structure that supports and cools two radii of pixel detectors.

This structure includes many of the generic features of future configurations; these include embedded cooling, high thermal conductivity inserts, engineered materials, and integrated electrical services.

### *4. R&D Directions*

There are a variety of R&D directions which could have significant impact on the thermal, mechanical, and electrical performance of support and cooling systems.

- Multiple Functions: Future R&D will continue to have mechanical structures perform multiple functions. For example there is ongoing work to make use of the electrical conductivity of carbon fiber facesheets to provide electrical shielding of sensors from high frequency clock signals as well as from fields generated from DC-DC converters.  R&D in this area would focus, for example, on material properties, co-curing, and new ways to integrate materials and services together.  Figure 9 shows and example from current R&D.

- Improved materials: This is a broad topic, central to the undertaking. It covers novel materials such as new ceramics, new carbon or other types of low density foams, adhesives, and high modulus fibers, among others.

- Engineered composites: The combination of existing and new materials into composite structures has already been fruitful. This R&D extends that line to new arrangements and configurations which could yield improved, or even novel, properties. Examples are laminate orientations and mixed materials.

- Materials science and nanotechnology: The materials science and nanotechnology community may offer approaches and materials which could have significant impact on R&D for future trackers. For example, recent work has indicated that very small percentage dispersions of nanomaterials in polymers can have dramatic effects on thermal conductivity and strength. This may lead to improved adhesives, and matrices for composite pre-pregs. Such preparations, and the processes need to create them, are state-of-the-art. It is important that our community engage where possible with chemists and materials scientists to further these developments.

- Spin-offs: As in the past, other fields may benefit from developments originating in HEP. Collaboration, in particular with the Basic Energy Sciences, Nuclear Physics, and Accelerator communities should be encouraged. New ideas can emerge from this interplay. Additional funding sources can sustain R&D capabilities through the inevitable funding cycles.

- Measurement techniques and QA: Characterization of materials and structures is also key to understanding their performance both in R&D and through fabrication. New and emerging tools should be considered, studied, and evaluated. Among these are 3D metrological methods, acoustical microscopy, large area automated inspection, thermal imaging, and precision survey, both static and dynamic.

- Simulation and Modeling: Analytical and numerical models of materials and composites are key tools in the development process. Recent work on finite element analysis of low density carbon foam has led to a predictive understanding of the dependence of thermal conductivity on density. Engagement with industrial and academic collaborators in this area should be encouraged.

- 3D printing: This is an emerging area which has received considerable attention. It has already been applied to create prototypical structures in a number of HEP applications. It should be watch as it develops further and in particular as new materials became available as printing media.

- Materials Database: Identifying and specifying materials of interest, and their properties, can sometimes be difficult to coordinate. For design, modeling, and simulation, it is important that workers use common and agreed upon values. A public access database of materials properties should be established. The database would include physical, structural, thermal, and radiation hardness properties. As time progresses, the database could be extended to include specific processes and techniques to match materials to low-mass detector requirements.

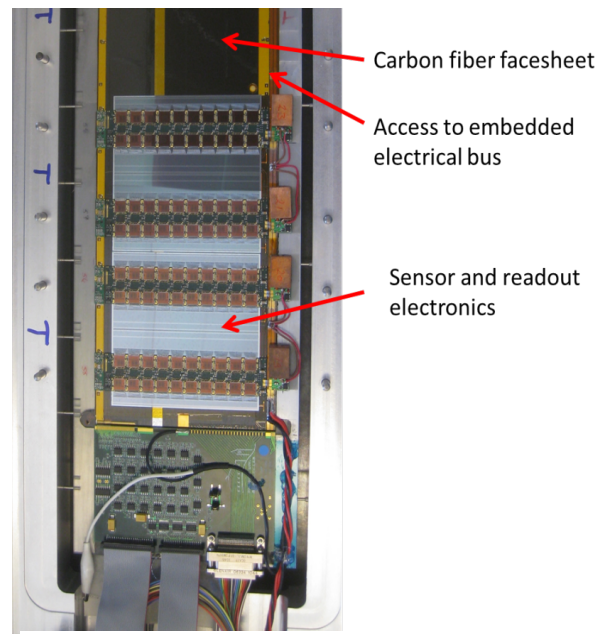

**Figure 9:** A current multiple function example. Sensor and readout electronics is mounted on structural laminate of carbon fiber facesheets, an embedded electrical bus cable, carbon foam and honeycomb core, and coolant carrying channels. The top face-sheet serves also as an electrical shield layer.

During the next ten years and beyond, a cooperative effort by industry, academic institutions, and national laboratories is essential to develop the materials, structures, tools, fabrication techniques, and thermal mechanical systems that will be required in next-generation and future detectors. Industrial involvement helps ensure that the latest materials and fabrication techniques are known and will be cost effective in appropriate quantities. Involvement of academic institutions and national laboratories is essential to monitor industrial developments for their applicability to physics needs, encourage crucial developments, and develop and test state-of-the-art new materials and processes. A cooperative effort among the three research and development communities helps minimize duplication of facilities and effort, thereby allowing development to proceed in a timely and cost-effective manner.

# Development of Large-Area Photosensors for Future HEP Experiments[14]

## *Introduction*

Photosensors provide essential functionality for HEP experiments in a wide variety of contexts. Increasingly, the demands of HEP experiments are pushing beyond standard capabilities of conventional phototubes. In this whitepaper, we examine the experimental needs of future HEP Experiments and how advances in photodetector technologies might impact their physics reach.

Photosensors are typically used in three broad categories of detector systems: (1) Large Cherenkov arrays and water Cherenkov tanks for full event reconstruction in neutrino experiments, proton decay, and detecting atmospheric showers from cosmic rays; (2) Ring Imaging Cherenkov (RICH) and Detectors of Internally Reflected Cherenkov light (DIRC) detectors used for particle identification in heavy flavor collider experiments; (3) and simple time-of-flight (TOF) systems used for vertex reconstruction and time-of-flight mass spectroscopy in colliders. The demands on these photosensors vary widely from one experiment to the next, but can be divided between the two main types of operational conditions:

1) Low-rate environments such as neutrino, double beta decay, and proton decay experiments. Under these conditions low-cost coverage, improved resolution for event reconstruction, and single photon counting capabilities are most important for background rejection, signal selection efficiency, tracking, and momentum resolution.

2) High-rate environments typical of collider based experiments. Here photodetectors need to be robust over years of operation under intense radiation, and capable of handling high event rates per unit area per unit time without saturating. Readout electronics must be designed to handle the high rates and track multiplicities.

Common to both experimental categories is the need to instrument large areas economically and the need for improvements in the combined timing and imaging resolutions.

## *Physics Needs and Desired Detector Capabilities*

Here we describe some of the areas ripe for improvements in photosensor technology and provide a few examples of physics issues to be addressed by these developments.

### *Cost per Area Coverage and Light Collection Efficiency*

Given the large scales of future collider (e.g. Atlas and CMS) and neutrino experiments (approaching megatons), a substantial improvement in the cost-per-area coverage of photodetectors can have a positive impact on large experiments. While this is not the only factor, improvements that allow better light collection and area coverage per unit cost can address key physics needs within the currently tight budgetary environment. For physics processes with low light yields, such as the detection of supernova neutrinos in large water Cherenkov detectors, better photon statistics translate directly to better energy resolution and improved sensitivity to the MeV-scale charged particles near the Cherenkov threshold [72]. The Daya Bay II collaboration provides another interesting example where light collection would enable new physics reach. In this experiment, a full spectrum analysis of the oscillation signal at many energies over the 60 km baseline would allow discrimination between the normal and inverted neutrino mass hierarchies. Thus the detectors for Daya Bay II would require excellent energy resolutions: 3

---

[14] Corresponding author(s) for this section: M. Wetstein[4], J. Va'vra[12], M.C. Sanchez[5]

MeV/√(E) compared with 8 MeV/√(E) in the comparable Kamland detector [73]. In the case of heavy flavor physics, the ability to replace multiple MCP-photosensors with single, large-area MCPs might also lower costs without compromising either time or imaging resolution.

For the same area coverage, light collection can be improved by advances in photocathode quantum efficiency. Not only the average quantum efficiency, but the spectral response of photocathodes is also of interest. Further gains can be made using light collectors, devices such as Winston Cones and wavelength shifting plastics. These allow for improved light collection with the same number of photosensors. Finally, in the context of large water Cherenkov experiments, it may be possible to employ full imaging optics such as parabolic and spherical reflective geometries. These approaches trade some angular phase space for increased area coverage. With the addition of precision timing to the spatial image, it may be possible to build a complete image, along with corrections for near field optical distortions from non-parallel incident light.

### *Imaging Capabilities*

Conventional PMTs are single pixel detectors. Position of the incident photons cannot be resolved except insofar as they hit a particular PMT. Photodetectors with spatial granularity would bring needed capabilities to HEP experiments. For example, such photosensors can be employed in the context of imaging optics, where the pattern of Cherenkov light is mapped to a smaller surface area by reflectors. In addition, Cherenkov- and scintillation-based neutrino detectors can make better use of the fiducial volume near the edge of the detector is it is possible to resolve photon positions within single photosensors. These capabilities are also important in RICH and DIRC systems, where the geometry and time-of-flight of the Cherenkov light is used for particle identification.

### *Precision Timing and Imaging Capabilities*

Many measurements in particle physics are limited by the time resolution of particle detectors. Time-of-flight systems approaching few picosecond resolutions would allow mass-spectroscopy and thus efficient flavor identification of high energy jets in collider experiments.

In many cases, the *combination* of precision timing and spatial resolutions is as important as the timing itself, allowing the measurement of event kinematics in terms of 4-vectors rather than 3-vectors. Measurements of the relative arrival times of particles with resolutions approaching single picoseconds over a drift distance of 1.5 m could be combined with track information to determine particle masses and thus the quark composition of hadronic jets. Timing measurements can also be used to match high energy gammas to their respective interaction vertices, providing a useful capability for collider experiments interested in measuring certain Higgs, KK, and gauge mediated SUSY signatures [74].

Similarly, in neutrino experiments where Cherenkov light is present, the combination of precision position and time information can be used to reconstruct tracks and EM showers with high resolution, which is important in reducing backgrounds for electron neutrino selection, as well as discrimination between gammas and electrons [75].

Heavy flavor physics experiments rely on technologies such as RICH (Ring Imaging Cherenkov) and DIRC (Direct Internally Reflected Cherenkov) detectors, where the geometric and timing properties of Cherenkov light emitted by high energy particles passing through thin radiators are used to separate between particles of similar momentum but differing mass. DIRC detectors, in particular, rely on timing to correct for chromatic dispersion effects on these Cherenkov rings. In all of these cases the ability to accurately measure both the positions and times of each arriving photon is very important [76].

### Single Photon Counting

In neutrino experiments, energy is largely measured by light yield. Accurate counting of individual photons in Cherenkov detectors is a necessary ingredient. Identification of single photons can be greatly improved by the combination of narrow pulse height distribution with fine-grained spatial and time separation. In low rate contexts, the ability to separate between photons in space and time enables essentially digital photon counting. In addition, this capability also enables improved separation between direct light and scattered light, prompt Cherenkov and scintillation light, and photons from multiple tracks. Precision photon counting could also improve the sensitivity of large Water Cherenkov detectors to low energy charged particles, close to the Cherenkov emission threshold, where observing small numbers of emitted photons above dark noise becomes challenging.

### Noise and Radiopurity

Photosensors are used in a variety of low-rate experiments. For deep underground neutrino experiments, dark rates from spontaneous firing of photodetectors -typically from the photocathode- reduce the signal-to-noise from Cherenkov and scintillation light [80]. The combination of coincidence triggers and time vetoes (in the case of beam neutrinos) can greatly ameliorate this effect. However, for rare, non-accelerator events with low light yields, dark rates can present a problem. The ability to make photosensors from very radio-pure materials is also an important capability for keeping backgrounds down in dark matter and double beta decay experiments [81].

### Robustness

Many HEP experiments subject photosensors to extreme conditions, particularly low temperatures, high pressures, and high radiation environments. Continued work on the implosion mechanisms stress-testing and containment are important for vacuum photosensors in deep water and ice applications. It is also important that photosensors can survive exposures to cryogenic temperatures such as those of liquid argon. Even functional aspects of the sensor, such as the resistivity of the photocathode and gain stages can depend on temperature and must be optimized for low-temperature operation.

### Radiation Hardness

In most collider-based environments, photosensors will be subject to unprecedented radiation doses. In Belle-II, detectors will see luminosities of $10^{36}$ cm$^2$ sec$^1$, with neutron doses up to ~$10^{11}$-$10^{12}$/cm$^2$, charged particle doses of ~$5 \times 10^{11}$/cm$^2$, and ~$5 \times 10^{11}$/cm$^2$ in just 10 years. In the central region of the Atlas Detector at 1 meter from the interaction point, a detector will see gammas and charged particles at a rate of ~$10^5$/cm$^2$, and neutrons at ~$10^6$/cm$^2$. These conditions require an R&D to develop radiation hard materials, such as fused silca and optical glues for DIRC-based detectors, SiPMT-based detectors with a good resistance to large neutron doses, photocathodes with a resistance to the radiation (a choice between bialkali and multi-alkali), electronics and many other materials. For example, SiPMTs develop a large noise rate and the pulse height resolution damage after a neutron dose of ~$10^{10}$/cm$^2$ [82].

### Rate Handling and Dynamic Range

Many high-rate experiments test the rate handling limits of photosensors. At RHIC and the LHC, the Star and Alice Experiments will see track multiplicities of 10,000 per event at a rate of 1kHz. The TORCH Experiment in LHCb will see a rate of roughly 10 MHz/cm2 and 10e11 photons per quartz detector per second [83], [84]. Optimizing photodetectors to operate at high rates without saturation or drop in gain is very important. In many contexts, particularly those of atmospheric Cherenkov arrays, photosensors must also be able to handle very wide ranges in light intensity [84].

## Recent and Ongoing Efforts

### Work with Conventional PMTs

The performance of large-area conventional phototubes has been studied in detail by a diversity of neutrino of potential experiments using large, monolithic, optical detectors [85].

As an example, recent design work by the Long Baseline Neutrino Experiment (LBNE) collaboration [85] targeting a 200-kton Water Cherenkov detector, resulted in the selection of the 12-inch Hamamatsu R11780 PMT. This is a 12" PMT available with peak quantum efficiency of 32% at 390 nm. The investigation of these tubes showed that they are good candidates for large optical detectors, due not only to their excellent performance but also a mechanical construction that can withstand high water pressures [82]. The performance measurements done showed peak-to-valley ratios greater than 2, transit time spreads around 1.3 ns and late-pulsing probabilities of less than 5% [82]. While the LBNE large Water Cherenkov design using 45,000 of these will not be going forward, the R&D is relevant to other possible optical detectors.

Similarly the HyperK working group [87] is working on the selection of phototubes for a 0.56 megaton fiducial Water Cherenkov detector, 10 times larger than the Super-Kamiokande detector. A fundamental requirement of this experiment is to be sensitive to a broad range of energies: from low MeV energies for supernova and solar neutrinos to atmospheric neutrinos in the GeV range. The baseline candidates for the Hyper-Kamiokande experiments are 99,000 large 20-inch PMT similar to those used in Super-Kamiokande (Hamamatsu R3600). Each of this would need to be surrounded by a protective case to avoid any accident cause by a chain-reaction starting from one phototube implosion. Other possibilities are of course being considered a 8 or 10-inch tube with a quantum efficiency of 40% which is almost double that of the R3600, the number of phototubes required would 1.4 times the number required for 20% coverage for the 20-inch models. Finally development is on-going on hybrid photodetectors (HPDs) which uses an avalanche photodiode instead of metal dynodes. These photodetectors provide high performance in timing and gain uniformity as well as fast response and better signal to noise ratio. Transit time spread for 1 PE have been measured to be approximately 0.6 ns for the 8-inch version and predicted to be 1.1 ns for the 20-inch version. These devices also demonstrate excellent photon separation with a peak to valley ratio of 8.9 compared to 1.9 in the conventional PMTs. The dark current in these devices might however be limiting for low energy physics in large water detectors.

Other collaborations such as LENA [86], Icecube [88], SNO [89] and MiniBoone [90] have done similarly detailed characterization studies of currently available photodetectors.

Work has also been done in designing elements that would enhance light collection efficiency such as Winston cones, wavelength shifting discs, and lensing systems. Such approaches can increase light collection up to 20-25%, usually at the cost of effective fiducial volume due to limited angles of acceptance, or decreased timing resolution [87], [91].

### LAPPD Collaboration

The Large Area Picosecond Photodetector (LAPPD) has developed flat-panel, large-area (8"x8") MCP-detectors for use in a variety of HEP contexts. These detectors consist of 8"x8" square MCPs made by thin-layer deposition of enhanced secondary electron emissive materials on substrates made from drawn glass capillary tubes. The vacuum package is based on flat-panel, sealed-glass technology, with no external pins. A single high voltage connection is made to the top window, which extends out of the vacuum region of the glass envelope. Operational voltages are set by a series of resistive spacers, and DC current through the MCP stack exits through the same stripline anode as the fast signal from MCP pulses. The delay-line, anode pattern consists of silver micro-striplines, silk-screened onto the bottom glass plate, and a frit-base vacuum seal is made over those striplines, between the detector wall and the bottom plate. The nominal LAPPD design addresses the entire detector system, including front-end electronics. The LAPPD collaboration has developed low-power ASICS, capable of sampling fast pulse shapes with 1.5 GHz analog bandwidth and up to 15 Msamples/sec sampling rates. Preliminary funding for this commercialization of the LAPPD design has been awarded through the DOE's STTR (Small business Tech Transfer) program [92]. Further reading on the LAPPD collaboration can be found later in this report.

### MLAP Collaboration

The Microchannel-Plate-Based Large Area Phototube Collaboration is a Chinese effort developing large spherical phototubes with small, 33mm MCPs as the gain stage. A pair of back-to-back MCP Chevrons share a single anode. Field lines from the photocathode focus photoelectrons on these MCPs. The microchannel plates are used as a gain stage, not for imaging purposes. Thus the MLAPC collaboration hopes to achieve reduced costs by taking MCPs which are adequate for amplification but flawed for use in image tubes. Another interesting aspect of the project is the combination of a transmission-mode and reflection-mode photocathode. Photons incident on the upper hemisphere may produce photoelectrons through the semitransparent cathode on top or they may pass through to the bottom layer, containing a thicker reflection mode photocathode. This is expected to increase the effective single-photon detection efficiency [93].

### Hybrid-APDs

This is a vacuum tube with enclosed solid state sensor of the avalanche photo-diode (APD). For example, in RICH application, Cherenkov photons enter through the entrance window and generate photoelectrons from a bialkali photocathode. Photoelectrons are accelerated in the electric field of 7–10 kV, and injected into the avalanche photodiode (APD). The APD provides an additional gain of ~40x. The Hamamatsu HAPD for the FRICH Belle-II detector consists of four APD chips, each of which is pixelated into 6x6 pads for a position measurement, and each pad has a size of 4.9mm x 4.9 mm. Small HAPD, for example a small Hamamatsu R10647U-01 with bialkali photocathode, have achieved an impressive single photon transit time spread of $\sigma \sim$ 28ps when illuminated over 8 mm dia. area, so they could be considered for some timing applications as well [94].

### SiPMTs (G-APDs or MMPCs)

These detectors could be used for RICH detector applications, especially their implementation in the array form. One of their weaknesses, especially for B-factory applications, is their sensitivity to neutron background. Presently, after a does of ~$10^{10}$ neutrons/cm$^2$, these detectors become very noisy. One has to also pay attention to temperature and voltage stability. The present timing resolution limit for a single 3mm x 3mm SiPMT with a 3cm-long quartz radiator is $\sigma \sim$ 16 ps/track. One area especially interesting is a development of digital SiPMTs, for example by Philips Company, where each SiPMT micro-pixel will carry information about timing and pulse height. This will lead to superbly percise detectors, both in position and timing [98].

### MRPCs (multi-gap RPCs)

For very large TOF systems one may consider MRPC technology, as pioneered by the ALICE detector. Its TOF resolution is close to $\sigma \sim$ 50ps/track, which is a great achievement, considering that the detector's total area ic close to ~150 m$^2$. It provides a good PID performance below ~1 GeV/c. With more gaps in the MRPC multi-electrode structure, one could push the timing resolution towards $\sigma \sim$ 10-20 ps/track. Similarly the STAR experiment reached a timing resolution of better than 100ps/track. Its total area was ~50m$^2$ [101].

### Ring Imaging and DIRC systems

The first focusing DIRC (FDIRC) was developed for the SuperB collider, where the photon camera would be outside of magnet. Its new focusing optics will make the photon camera ~25x smaller than the BaBar DIRC, while it timing resolution with H-8500 MaPMTs will be ~10x better. The photon camera is outside of the magnetic field, which simplifies the detector requirements. The concept relies on a pixel-based reconstruction of the Chrenkov angle; timing is used only to reduce the background, reduce some ambiguities and do chromatic corrections. It would have a similar performance as the BaBar DIRC, i.e., a 3σ π/K separation at ~4 GeV/c. One should note that the present FDIRC camera needs 48 H-8500 PaPMTs, however, it would need only 4 LAPPD 8"x8" detectors. Since the SuperB was cancelled, this detector is now one option for the future experiment at Electron-Ion collider at either BNL or JLAB, if one wants to re-use BaBar quartz radiator bar boxes.

One should also mention a TOP counter development for Belle-II, which relies heavily on the high resolution timing at a level of 50-100ps to make the Cherenkov angle reconstruction, and much less on the pixel information. The MCP-PMT detectors have to operate in the magnetic field. This detector will also pioneer a waveform digitizing electronics.

There is a new effort to develop FDIRC detectors for Panda experiment and future experiments at Electron-Ion colliders. These detectors employ various new focusing optics schemes trying improve the Cherenkov angle resolution and achieve a 3σ π/K separation up to ~6 GeV/c. This requires to improve both the single photon Cherenkov angle resolution and a total number of measured Cherenkov photons to achieve a resolution of $\sigma_{track}$ ~ 1-1.5 mrads (BaBar DIRC achieved $\sigma_{track}$ ~ 2.5 mrads). For photon detectors they are investigating either the SiPMT-array readout or new MCP-PMTs with 3-5 micron holes trying to operate up to ~3 Tesla (for a solution when the photon camera is inside the magnet). There is a vigorous R&D program both in simulation and hardware areas. This is certainly a new domain in DIRC-like detector development, if successful.

However, for very high momenta PID applications one has to use a gaseous radiator. An example is the forward RICH R&D development for the Electron-Ion collider. It uses a CsI photocathode with multi-GEM photon detector readout.

Another class of RICH detectors worth of mentioning is based on multi-layer Aerogel radiator, coupled to, for example MCP-PMTs or HAPDs. The refraction indices of different Aerogel layers are chosen so one achieves a focusing of the Cherenkov image. The Belle-II plans to use HAPDs, SuperB was considering MCP-PMTs. SuperB detector could achieve a 3σ π/K separation at ~5 GeV/c, which is better that FDIRC barrel performance, which aimed for 3σ π/K separation at ~4 GeV/c. The SuperB forward RICH detector proposal assumed 550 Planacon MCP-PMTs, but it would take only twelve 8"x8" LAPPD MCP-PMTs, if they would be available [104].

### Simple TOF systems

The most simple TOF concept, as it as it avoids complicated 3D data analysis and minimizes chromatic effects, is based on MCP-PMT detector coupled to polished and side-coated fused silica radiator cubes, which are optically isolated from each other. It requires a large number of MCP-PMT detectors, which increases the cost prohibitively at present. Although the prototype beam test reached σ ~ 14 ps/track, the cost was the reason why SuperB did not choose this type of PID concept. However, if MCP-PMT detectors would become cheap in future, one could revive this concept again because it is simple. For example, it may become possible with LAPPD MCP-PMT detectors [74], [109].

### DIRC-like TOF systems

The early TOF detectors measured time only and typically used scintillator as a radiator. The recent and future TOF detectors measure not only time but x & y coordinates as well, and use a precise quartz radiator, so that a trajectory of each photon can be calculated precisely. The time is constrained by the Cherenkov light information. They are typically called DIRC-like TOF detectors. Examples are (a) TORCH DIRC detector at LHCb, (b) Panda Disc DIRC detector, or (c) FTOF DIRC-like detector for SuperB, etc. The TORCH detector is expected to achieve σ ~15ps/track, Panda Disc detector is expected to reach σ ~50-100 ps/track, and FTOF DIRC-like detector for SuperB was supposed to reach σ ~70 ps/track [114].

### *MCP-PMT and SiPMT rate capability and lifetime for B-factories*

As an example, the expected rate of H-8500 MaPMT in FDIRC at SuperB would be about $2\text{-}3\times10^4$ counts/cm$^2$/sec and it would expected to reach ~0.2 C/cm$^2$/10 years of operation after integrated luminosity of ~50ab$^{-1}$. Belle-II TOP counter detector, because one deals with smaller detector area, will reach ~0.8 C/cm$^2$/10 years of operation. The bench tests show that most MCP-PMTs show a stable single photon operation up to 200-300 kHz/cm$^2$ at gain of $10^6$. Some MCP-PMTs, for example Hamamastu R10754-00-L4 with 10 micron holes, can reach ~5 MHz/cm$^2$. This means that in both above examples, detectors will be able to handle the rate. However, a possible problem is with the photocathode quantum efficiency (QE). In one case, one expects that the H-8500 MaPMT has a normal PMT aging behavior, and would be fine with the SuperB. However, in the second case, according to Belle-II studies, the best MCP-PMTs can safely reach 0.8-1 C/cm$^2$ before the QE start showing some degree of deterioration, and this is somewhat close to expected a total dose after integrated luminosity of ~50ab$^{-1}$. More studies are needed [116].

# Solid State Photodetectors[15]

The conversion of optical signals to electronic signals has been a central feature of high energy physics experimental techniques since the very early days. Initially photomuliplier tubes were the principal method used for this purpose. With the advent of the semiconductor industry silicon photodiodes, CCDs and Avalanche Photodiodes (APDs) have become common tools in HEP experiments. Very low-noise CCDs have been used in astronomy and HEP groups have contributed significantly to their application into the infra-red to enable the detection of the high-*z* galaxies. In HEP experiments the challenge of reading out large numbers of channels at a low cost has been met with solid state photodetectors. The introduction of new photodetector techniques to HEP has tracked progress in industry, in most case preceding common adoption by industry.

In the past few years there has been a revolution in our methods of light detection with the introduction of pixelated avalanche photodiodes operating in the Geiger mode with passive or active quenching circuits. These silicon devices are most commonly known as SIPMs or MPPCs. When a photon strikes a SiPM the pixel goes into Geiger mode, saturates and generates a large current. The sum of the currents from all the struck pixels is added to produce a signal which is linear to the flux of photons impinging on the device. After an interval determined by the device design the pixel is brought back to its original state and is fully efficient. As these devices are made using standard CMOS silicon processing techniques they are relatively cheap to manufacture and have been adopted in many applications both inside and outside HEP. One notable property is that their excess noise factor is small (~ 1.3), comparable to the best PMTs. This is unlike APDs operating in linear mode where the noise increases linearly with the device gain.

The great advantage of these SiPMs is that they are silicon based, behind which lays a large industry that produces material of extraordinarily high quality and with fabrication facilities found around the world. The disadvantage for some applications in HEP is that they are made with silicon. Thus in some applications where the radiation levels are very high the dark current produced by lattice damage can introduce high levels of background into the detector. To mitigate this it is possible to cool the devices, where one can achieve a factor of two reductions in dark count for every interval of 7 $^{\circ}$C that the device is cooled. However for some real-world applications this can be too difficult a drawback. Another drawback to these devices is that silicon has a very short attenuation length for light in the near UV. This makes their use for the detection of scintillation light from noble liquids difficult (for example Xe or Ar for the detection of dark matter). To circumvent this, wavelength shifters are used, with the concomitant loss in photon detection efficiency.

Because of their relatively low cost, silicon avalanche photodetectors have been used extensively in the field. The CMS lead tungstate calorimeter is readout with a total of 120,000 linear mode APDs and the whole of the NOvA detector is to be readout with APDs. Some of the applications of SiPMs, actual or planned, are the T2K fine grained detector, the upgraded CMS hadronic calorimeter. Outside of the HEP their use in PET detectors is becoming standard.

For future detectors that are being considered by the field they are certain to have many useful applications, especially where a low cost readout of a large number of channels is required. Nevertheless, the limitations of silicon based devices leads to consideration of alternative materials for photodetectors. These can be considered in terms of wavelength of the light to be detected. Short wavelength detection of light is a central problem in the detection of Cherenkov radiation and silicon is not well adapted to this. The detection of UV light, especially for wavelengths less than 360 nm, is particularly difficult for silicon based devices. This is an area of active research in industry for use in monitoring jet engines and for well logging and there is an active area of research in (4H:SiC) Silicon Carbide, which has a bandgap of 3.23 eV.

---

[15] Corresponding author(s) for this section: R. Rusack[13]

In industry one of the main interests is the detection of light in the near infrared, specifically at wavelengths of 1300 nm and 1550 nm for the detection of modulated light in data communications. Silicon photodetectors are not sensitive in this region due to their bandgap of 1.12 eV, corresponding to a wavelength cut off of 1100 nm. Interesting detectors where Germanium, which has a bandgap of 0.66 eV, are being investigated. The Ge is grown on a silicon substrate, which maintains the advantages of integrated electronics of silicon, combined with the infra-red detection capabilities of germanium.

In another recent development multi-pixel photo-counters based on GaAs and InGaAs, have been reported. These devices should be intrinsically more radiation hard than silicon and may find a place for their application in calorimeters where the radiation levels are higher than can be sustained by silicon devices.

Specific properties that are likely to be relevant for semiconductor photodetectors, apart from the generic low cost, by application are as follows.

*Ultra high-energy cosmic ray detection of Cherenkov radiation:*

UV sensitivity, fast signal response.

*Calorimeters:*

HL-LHC: radiation tolerance, large gain band-width product, large dynamic range

ILC: very low cost, large gain-bandwidth product.

*High redshift galaxy detection:*

Extended sensitivity into the infra-red.

# Crystal Detectors[16]

Crystal detectors have been used widely in high energy and nuclear physics experiments, medical instruments, and homeland security applications. Novel crystal detectors are continuously being discovered and developed in the academy and in industry. In high energy and nuclear physics experiments, total absorption electromagnetic calorimeters (ECAL) made of inorganic crystals have been known for decades for their superb energy resolution and detection efficiency for photon and electron measurements. Crystal ECAL is the choice for those experiments where precision measurements of photons and electrons are crucial for their physics missions. Examples are the Crystal Ball NaI(Tl) ECAL, the L3 BGO ECAL and the BaBar CsI(Tl) ECAL in lepton colliders, the kTeV CsI ECAL and the CMS PWO ECAL in hadron colliders and the Fermi CsI(Tl) ECAL in space. For future HEP experiments at the energy and intensity frontiers, however, the crystal detectors used in the above mentioned ECALs are either not bright and fast enough, or not radiation hard enough. Crystal detectors have also been proposed to build a Homogeneous Hadron Calorimeter (HHCAL) to achieve unprecedented jet mass resolution by duel readout of both Cherenkov and scintillation light, where the cost of crystal detectors is a crucial issue because of the huge crystal volume required. This section discusses several R&D directions for the next generation of crystal detectors for future HEP experiments.

## *Crystal detectors with excellent radiation hardness:*

With $10^{35}$ cm$^{-2}$s$^{-1}$ luminosity and 3,000 fb$^{-1}$ integrated luminosity the HL-LHC presents the most severe radiation environment, where up to 10,000 rad/h and $10^{15}$ hadrons/cm$^2$ are expected by the CMS forward calorimeter located at pseudorapidity $\eta$=3. Superb radiation hardness is required for any detector to be operationally stable in the HL-LHC environment. Fast timing is also required to reduce the pile-up effect. Cerium doped lutetium oxyorthosilicate (Lu$_2$SiO$_5$ or LSO) and lutetium-yttrium oxyorthosilicate (Lu$_{2(1-x)}$Y$_{2x}$SiO$_5$ or LYSO) are widely used in the medical industry and have attracted much interest in the HEP community. With high density (7.1 g/cm$^3$), fast decay time (40 ns) and superb radiation hardness against gamma-rays and charged hadrons LSO/LYSO crystals are good crystal detectors for future HEP experiments at both the energy and intensity frontiers. LSO/LYSO crystals have been chosen by the Mu2e experiment. A LSO/LYSO crystal based sampling ECAL has also been proposed for the CMS forward calorimeter upgrade for the HL-LHC. Because of its high density, short light path (as compared to the total absorption ECAL) and excellent radiation hardness a LSO/LYSO crystal based sampling ECAL is expected to provide a compact and stable calorimeter at the HL-LHC.

R&D issues for crystal detectors with excellent radiation hardness include the following aspects:

- While the gamma-ray and proton induced radiation damage in crystal detectors has been investigated, only limited data are available for neutron induced damage in scintillation crystals. An R&D program to investigate radiation damage effects induced by gamma-rays, charged hadrons as well as neutrons, in various crystal detectors is needed to address this issue.

- LSO/LYSO crystals are relatively expensive because of the high cost of the raw material, lutetium oxide, and its high melting point. Other fast scintillators (See Table 8 below) as well as fast scintillating ceramics may also be considered. Development of fast and radiation hard crystal and ceramic detectors is of general interest.

- For an LSO/LYSO based Shashlik ECAL, radiation hard photo-detectors as well as wavelength shifters need also to be developed.

---

[16] Corresponding author(s) for this section: R-Y. Zhu[14]

## Crystal detectors with fast response time:

Future HEP experiments at the energy frontier, such as HL-LHC, and the intensity frontier, such as the proposed project X, require very fast crystals to cope with the unprecedented event rate.

| | LSO/LYSO | GSO | YSO | CsI | BaF$_2$ | CeF$_3$ | CeBr$_3$ [1] | LaCl$_3$ | LaBr$_3$ | Plastic scintillator (BC 404) [2] |
|---|---|---|---|---|---|---|---|---|---|---|
| Density (g/cm$^3$) | 7.40 | 6.71 | 4.44 | 4.51 | 4.89 | 6.16 | 5.23 | 3.86 | 5.29 | 1.03 |
| Melting point (°C) | 2050 | 1950 | 1980 | 621 | 1280 | 1460 | 722 | 858 | 783 | 70[#] |
| Radiation Length (cm) | 1.14 | 1.38 | 3.11 | 1.86 | 2.03 | 1.70 | 1.96 | 2.81 | 1.88 | 42.54 |
| Molière Radius (cm) | 2.07 | 2.23 | 2.93 | 3.57 | 3.10 | 2.41 | 2.97 | 3.71 | 2.85 | 9.59 |
| Interaction Length (cm) | 20.9 | 22.2 | 27.9 | 39.3 | 30.7 | 23.2 | 31.5 | 37.6 | 30.4 | 78.8 |
| Z value | 64.8 | 57.9 | 33.3 | 54.0 | 51.6 | 50.8 | 45.6 | 47.3 | 45.6 | - |
| dE/dX (MeV/cm) | 9.55 | 8.88 | 6.56 | 5.56 | 6.52 | 8.42 | 6.65 | 5.27 | 6.90 | 2.02 |
| Emission Peak[a] (nm) | 420 | 430 | 420 | 420 / 310 | 300 / 220 | 340 / 300 | 371 | 335 | 356 | 408 |
| Refractive Index[b] | 1.82 | 1.85 | 1.80 | 1.95 | 1.50 | 1.62 | 1.9 | 1.9 | 1.9 | 1.58 |
| Relative Light Yield[a,c] | 100 | 45 | 76 | 4.2 / 1.3 | 42 / 4.8 | 8.6 | 141 | 15 / 49 | 153 | 35 |
| Decay Time[a] (ns) | 40 | 73 | 60 | 30 / 6 | 650 / 0.9 | 30 | 17 | 570 / 24 | 20 | 1.8 |
| d(LY)/dT[a,d] (%/°C) | -0.2 | -0.4 | -0.1 | -1.4 | -1.9 / 0.1 | ~0 | -0.1 | 0.1 | 0.2 | ~0 |

a. Top line: slow component, bottom line: fast component.
b. At the wavelength of the emission maximum.
c. Relative light yield normalized to the light yield of LSO.
d. At room temperature (20°C)
#. Softening point

1. W. Drozdowski et al. *IEEE TRANS. NUCL. SCI*, VOL.55, NO.3 (2008) 1391-1396
   Chenliang Li et al, *Solid State Commun*, Volume 144, Issues 5–6 (2007),220–224
   http://scintillator.lbl.gov/
2. http://www.detectors.saint-gobain.com/Plastic-Scintillator.aspx
   http://pdg.lbl.gov/2008/AtomicNuclearProperties/HTML_PAGES/216.html

**Table 8:** Basic Properties of Fast Crystal Scintillators

Table 8 above summarizes basic properties of fast crystal scintillators, including oxides (LSO/LYSO, GSO and YSO), halides (BaF$_2$, CsI and CeF$_3$) and recently discovered bright and fast halides (CeBr$_3$, LaCl$_3$ and LaBr$_3$), and compares them to typical plastic scintillator (last column). All fast halides discovered recently are highly hygroscopic so that their application needs extra engineering work. Among the non-hygroscopic crystal scintillators listed in Table 8, BaF$_2$ crystals are distinguished with their sub-nanosecond fast decay component, that produces high scintillation photon yield in the 1$^{st}$ nano-second, just less than highly hygroscopic CeBr$_3$ and LaBr$_3$. This fast component provides a solid foundation for an extremely fast ECAL made of BaF$_2$ crystals. In addition to the good energy resolution and fast response time the photon direction measurements are also important for neutral pion identification through its two photon decays. Most crystal ECAL built so far has no longitudinal segmentation because of the technical difficulty of imbedding readout device in a total absorption calorimeter. One exception is the Fermi ECAL which is a CsI(Tl) strip based hodoscope. With compact readout devices developed in the last decade this issue is less relevant now. A well designed longitudinally segmented ECAL with fast crystal detectors may provide excellent energy resolution, fast timing and fine photon angular resolution, so would serve well for future HEP experiments in general and those at the intensity frontier in particular.

R&D issues for crystal detectors with very fast response time include the following aspects:

- The slow scintillation component in BaF$_2$ needs to be suppressed by selective doping and/or selective readout by using solar blind photo-detector. The radiation harness of BaF$_2$ crystals needs to be further improved.

- Development of novel fast crystal detectors with sub-nanosecond decay time, e.g. CuI etc., is of general interest.
- For a longitudinally segmented crystal calorimeter with photon pointing resolution, R&D on detector design and compact readout devices are required.

### *Crystal detectors for excellent jet mass reconstruction:*

The next generation of collider detectors will emphasize precision for all sub-detector systems. One of the benchmarks is to distinguish W and Z bosons in their hadronic decay mode. Excellent jet mass resolution is required for future lepton colliders at the energy frontier. Following successful experiences with crystal ECAL, a novel HHCAL detector concept was proposed to achieve unprecedented jet mass resolution by duel readout of both Cherenkov and scintillation light. The HHCAL detector concept includes both electromagnetic and hadronic parts, with separate readout for the Cherenkov and scintillation light and using their correlation to obtain superior hadronic energy resolution. It is a total absorption device, so its energy resolution is not limited by sampling fluctuations. It also has no structural boundary between the ECAL and HCAL, so it does not suffer from the effect of dead material in the middle of hadronic showers. With the dual-readout approach large fluctuations in the determination of the electromagnetic fraction of a hadronic shower can be reduced. The missing nuclear binding energy can then be corrected shower by shower, resulting in a good energy resolution for hadronic jets. Such a calorimeter would thus provide excellent energy resolution for photons, electrons and hadronic jets.

R&D issues for crystal detectors for the HHCAL detector concept include the following aspects:

- Development of cost-effective crystal detectors is crucial because of the unprecedented volume (70 to 100 m$^3$) foreseen for an HHCAL. The material of choice must be dense, cost-effective, UV transparent (for effective collection of the Cherenkov light) and allow for a clear discrimination between the Cherenkov and scintillation light. The preferred scintillation light thus is at a longer wavelength, and not necessarily bright or fast.
- The HHCAL requires R&D on detector design and on compact readout devices.

| Parameters | $Bi_4Ge_3O_{12}$ (BGO) | $PbWO_4$ (PWO) | $PbF_2$ | PbClF | $Bi_4Si_3O_{12}$ (BSO) |
|---|---|---|---|---|---|
| ρ (g/cm$^3$) | 7.13 | 8.29 | 7.77 | 7.11 | 6.8 |
| $λ_I$ (cm) | 22.8 | 20.7 | 21.0 | 24.3 | 23.1 |
| n @ $λ_{max}$ | 2.15 | 2.20 | 1.82 | 2.15 | 2.06 |
| $τ_{decay}$ (ns) | 300 | 30/10 | ? | 30 | 100 |
| $λ_{max}$ (nm) | 480 | 425/420 | ? | 420 | 470 |
| Cut-off λ (nm) | 310 | 350 | 250 | 280 | 300 |
| Light Output (%) | 100 | 1.4/0.37 | ? | 17 | 20 |
| Melting point (°C) | 1050 | 1123 | 842 | 608 | 1030 |
| Raw Material Cost (%) | 100 | 49 | 29 | 29 | 47 |

**Table 9:** Basic properties of candidate crystal detectors for HHCAL

Table 9 above compares basic properties of candidate cost-effective crystal detectors for the HHCAL detector concept, such as PWO, scintillating $PbF_2$, PbFCl and BSO. Scintillating glasses and ceramics are also attractive for this detector concept.

## *Summary*

Crystal detectors have been used widely for decades in high energy and nuclear physics experiments, medical instruments and homeland security applications. Future HEP experiments require bright and fast crystal detectors with excellent radiation hardness. Cost-effectiveness is also a crucial issue for crystal detectors to be used in a large volume. Crystal detectors are also an area of fast development. New materials are continuously being found each year. To face these challenges a thorough R&D program is required to investigate and develop crystal detectors for future HEP experiments in all frontiers.

# The Large Area Picosecond Photodetector (LAPPD) Project[17]

For decades, the high energy physics (HEP) community has relied on photomultiplier tubes (PMTs) to provide low cost, large-area coverage for a wide variety of detector systems. Increasingly, the demands of HEP experiments are pushing for new imaging capabilities, combined with temporal resolutions far better than PMTs can typically offer. A possible replacement for the PMT is the microchannel plate photomultiplier tube (MCP-PMT) [117], a compact detector capable of micron-level spatial imaging and with single photoelectron time resolutions typically measured in tens of picoseconds [118].

The Large Area Picosecond Photodetector (LAPPD) Collaboration [122] is developing large-area, flat panel MCP detectors. Central to the LAPPD project is the use of Atomic Layer Deposition (ALD) to mass-produce MCPs from low cost glass substrates. ALD is a batch process whereby materials can be applied uniformly and conformally to large surface areas in bulk, one molecular mono-layer at a time [123]. The structure of LAPPD-made MCPs is provided by borosilicate glass substrates, cut from hexagonally-packed bundles of drawn capillaries with 20-micron pore structures and open-area ratios of ~70%. The substrates are ALD coated, first with a layer of resistive material and then with a secondary-emitting layer. This approach allows independent optimization of the geometric, resistive, and secondary-emission characteristics of MCPs [124].

The complete LAPPD MCP detector comprises a photocathode, the MCPs, a segmented anode, and high voltage distribution, integrated within a sealed, flat-panel vacuum envelope. Signals from the MCPs are received by a microstripline anode structure, optimized for high-bandwidth electronics [127]. This delay-line design greatly reduces the necessary channel count, as electronic resources scale only with the square root of the area. Hit positions are determined by the signal centroid in the direction perpendicular to the striplines and, along the strips, by the difference in the arrival time at the two ends of the striplines. The digitization electronics uses low-power CMOS technology [128]. Arrival times and gains of the pulse trains are measured by waveform sampling, which offers pulse shape and charge measurement, multiple hit capability, as well as the best timing resolution [129]. The entire package is hermetically sealed in a flat-panel glass body [130].

Tests of working LAPPD detector systems have demonstrated 8" MCPs with gains above $10^7$. Time resolutions approach single picoseconds in the large signal limit (N photoelectrons > 100) and better than 50 picoseconds for single photoelectrons. The position of incident photons in the direction parallel to the stripline anode is determined by the difference in signal arrival time at the two ends of our delay line. This differential time resolution was measured to be better than 20 psec for single photoelectrons, which would translate to a spatial resolution of roughly 2 mm in the direction parallel to the strips. Position in the transverse direction is determined by a weighted centroid of the charge collected on each strip and is measured to be better than 1 mm [131].

Nominally LAPPDs will be made with conventional photocathodes. The collaboration has been able to demonstrate 8" x 8" photocathodes with quantum efficiency above 20% [132]. As these detectors are commercialized, it is safe to assume that they can be made with QE's comparable to standard phototubes available on the market.

Another major milestone of the project was the production of the PSEC4 ASIC, a fast sampling chip capable of digitizing MCP pulses at 15 giga-Samples/second, with an analog bandwidth of better than 1 GHz and less than 1 mV noise [128]. The ASIC is made using low-power and scalable CMOS technology. Along with the PSEC4, a full front-end FPGA-based DAQ system is now being tested. The near-complete

---

[17] Corresponding author(s) for this section: H. Frisch[15], M. Wetstein[4] on behalf of the LAPPD Collaboration

detector system is now being tested on a "demountable LAPPD", a working seal-glass tube, whose only differences from the final design goal are that: 1) the tube is actively pumped rather than hermetically sealed; 2) the seal between the top window and the tube body is with an O-ring rather than an indium seal; and 3) the photocathode is a thin Aluminum layer rather than a bialkali, as the demountable is assembled in air. The purpose of the demountable LAPPD setup is to test a complete, end-to-end detector system, including our own LAPPD-designed front-end electronics. The system allows readout of all 60 detector channels, 30 per side. A four anode chain constitutes one row of three in a "Super Module" (SuMo) - a large-area detector system designed to reduce channel counts by sharing the same delay line pattern for several MCPs [129].

The Department of Energy awarded a Phase I STTR grant (Section 38) to Incom Inc. for technology transfer for LAPPD detectors [133]. At Berkeley Space Science Laboratory a parallel effort is underway to produce MCPs with an alternative ceramic-body design.

LAPPDs may prove useful in a wide variety of non-HEP contexts, such as medical imaging, neutron detection, and homeland security. The HEP community could greatly benefit from the economies of scale achievable with a larger market.

# Transition Edge Sensors for High Energy Physics[18]

## *Transition Edge Sensors*

Transition Edge Sensors (TES) are ultra-sensitive thermometers consisting of a thin superconducting film that is weakly heat-sunk to a bath temperature much lower than the superconductor Tc (see Figure 10, left). The principles of operation are simple to understand. By supplying electrical power to the TES, we can raise the temperature of the sensor so that the film is in the middle of its superconducting-to-normal transition (see Figure 10, right). If the electrical power is supplied via a voltage bias, a negative feedback loop is established. Small changes to the TES temperature, arising from thermal fluctuations (noise) or changes in the absorbed power from a source (signal), lead to large changes in the TES resistance. The change in resistance creates a canceling effect because increases (or decreases) in temperature produce decreases (or increases) in Joule heating power. This negative electro-thermal feedback is very strong because the transition is very sharp. It linearizes the detector response and expands the detector bandwidth. The fundamental noise arises from thermal fluctuations. At the sub-K operating temperatures for typical TES devices, these fluctuations are small giving the TES excellent resolving power. An extremely important aspect of TES devices is that they are low-impedance (<1 Ω) and can be multiplexed with modern-day Superconducting QUantum Interference Device (SQUID) multiplexers. This makes TES-based detectors amenable to large array applications.

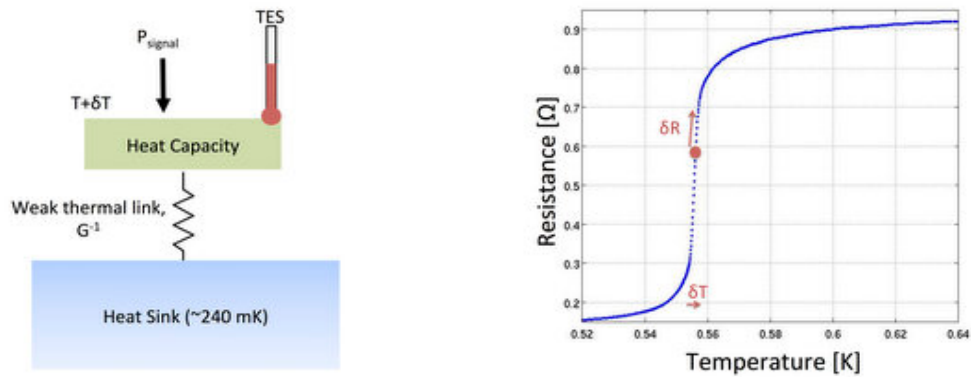

**Figure 10:** Left: Illustration of a thermal circuit for a typical Transition Edge Sensor (TES) detector highlighting the principles of signal detection. A weakly thermally sunk heat capacity absorbs power, Psignal, which is to be measured. Variations in the absorbed power change the heat capacity's temperature, which is measured by a TES operating under strong electro-thermal feedback. Right: Plot of resistance versus temperature for a typical TES illustrating the principles of negative electro-thermal feedback. The TES is voltage biased into the middle of its superconducting-to- normal transition. Small changes in the TES temperature produce large changes in the TES resistance. Since the TES is voltage biased, an increase (or decrease) in the temperature produces an increase (or decrease) in the resistance leading to a decrease (or increase) in the Joule heating power supplied by the bias. This canceling effect corresponds to a strong negative electro-thermal feedback making the current through the TES nearly proportional to Psignal.

---

[18] Corresponding author(s) for this section: C.L. Chang[4], B. Caberera[16], J.E. Carlstrom[4], S. Golwala[14], K.D. Irwin[16], A.T. Lee[6], M. Pyle[17]

### TES Detectors and the Cosmic Frontier

TES detectors were invented within the context of Dark Matter searches. The development of TES detectors with exceptional background rejection has pioneered the field of direct Dark Matter detection. The strong detector history has provided a solid understanding of the fundamental detector physics along with a proven track record of scientific leadership. Over the course of the next decade, TES-based Dark Matter detectors continue to be a promising technology. There is a clear path towards a G3 ton-scale Dark Matter observatory and on-going R&D focused on reducing the detector energy threshold which would increase the sensitivity of TES Dark Matter detectors to lower mass WIMPs.

Over the last decade, TES bolometers have also had a profound impact on CMB measurement. CMB detector technology is sensitivity limited where the measurement is fundamentally limited by the shot noise of the photons being measured. As such, increases in the sensitivity of experiments is possible only through increasing the size of the focal plane. At the end of the last century, CMB detector technology resembled that of infrared imaging at its inception: limited hybrid focal plane arrays involving a small number of individual detectors. The implementation of the HEP invented TES to CMB bolometers together with modest multiplexing has enabled kilo-pixel focal plane arrays which have ushered in an era of unprecedented CMB surveys. The next decade of CMB measurement and instrumentation development is focused on Stage III (>10,000 detector elements) and Stage IV (>100,000 detector elements) experiments with each stage having an order of magnitude increase in experiment mapping speed. The Stage III and Stage IV sensitivities translate directly into the size of fielded focal planes and TES-based focal planes are the only CMB detector technology with a clear path to Stage IV sensitivities.

### Challenges for TES detectors R&D

The technology development challenges for the next decade of TES detectors are similar across both applications within the HEP Cosmic Frontier and are broken down as follows:

- **Extending the reach of active detector elements** For Dark Matter TES detectors, this challenge involves reducing the detector noise by nearly an order of magnitude. Such a reduction would lower the detection threshold by a factor of ten opening up sensitivity to new parameter space for low mass WIMPs. For CMB bolometers, this challenge involves developing reliable fabrication of high bandwidth ultra-low loss superconducting microstrip which is utilized for coupling the bolometer to mm-wave radiation.

- **Large increases in the volume and throughput of detector production** Upcoming TES applications for both Dark Matter and CMB experiments require dramatic increases in the fabrication of TES detectors. In order to achieve the required production throughput, access to micro-fabrication resources is required with exclusive control of the thin film deposition systems. Together with developing the fabrication processes and tooling for mass production of TES devices, an extensive program of detector testing, characterization, and quality control is necessary to support the increased production capacity. Additionally, improvements in the multiplexed readout of TES detectors is required for high-bandwidth operation of future TES detector arrays with thousands of elements.

The US leads the world in the implementation of TES technology to the forefront of HEP science. TES detectors have driven the field of Dark Matter detection and are the best technology for ongoing leadership in CMB experiment. The technical challenges for the next decade of TES development reflect the fact that the superconducting technology critical for future Dark Matter and CMB science is not available through commercial industry. As such, HEP resources and facilities are required to sustain ongoing US leadership in these fields.

## *Broader Impact*

TES technology is exceptionally versatile with diverse applications beyond those listed above. Over the last decade TES detectors have been implemented in the following:

- measurements of the neutrino mass via both double beta decay and beta decay endpoint measurements
- quantum information and cryptography
- spectrophotometry for optical astronomy
- focal planes for sub-mm astronomy
- detectors for national security include thermal imaging and nuclear non-proliferation
- micro-calorimeter arrays x-ray astrophysics
- spectroscopy at synchrotron light sources

The spectrum of TES applications illustrate how ongoing HEP investment into TES development can produce technologies that are readily exported to programs and disciplines beyond the scope of HEP science.

# Instrumentation Challenges for Direct Detection of Dark Matter[19]

In this section we review dark matter sensor technologies exclusive of the liquid noble gas systems. Common to all of these sensors, and the liquid noble gas detectors, is the overarching challenge of ultra-pure materials and assay techniques, covered at the end of this contribution.

## *HPGe Spectrometers*

Current and planned experiments using high-purity germanium (HPGe) spectrometers include CoGeNT, C4, MAJORANA, and GERDA. While the MAJORANA and GERDA experiments are primarily searches for neutrinoless double-beta decay, if they can achieve the required low energy threshold - even on a fraction of their detector mass - they will be formidable DM detectors. The main challenges for these experiments are background reduction (see materials and assay section) and lowering energy thresholds to <100 eV.

CoGeNT/C4 employs commercially-available PPC (p-type point contact) germanium crystals to drastically reduce the capacitance of the detector element, and with it the contributions from several sources of electronic noise. This results into a present energy threshold of 0.45 keVee (~2 keVnr), an order of magnitude improvement over conventional coaxial germanium detectors [134]. The technology is inexpensive and stable enough for a demonstrated continuous operation spanning several years [135], i.e., highly suitable for a search for the annual modulation signature from low-mass WIMPs, testing the DAMA/LIBRA anomaly. For 1 pF capacitances like those obtained in these PPCs, the theoretically reachable electronic noise is however one full order of magnitude lower (around 50 eVee, commonly achieved in state-of-the-art x-ray semiconductor detectors). The dominant source of residual noise in past CoGeNT detectors has been identified as parasitic capacitances affecting the first stage of signal amplification. Measures against this source of noise have been implemented in the PNNL-built cryostat for the first C4 detector [cogent3], about to be tested at the time of this writing. Beyond those, the manufacturer of these PPCs (Canberra Industries) has embarked in studies of further optimization of intra-contact surface passivation and point-contact design, both able to generate a further noise reduction. A team at the University of Washington works on a new design of pulse-reset preamplifier, again able to improve the noise budget. In addition to these efforts, the University of Chicago group is developing FPGA-based intelligent triggering schemes capable to reduce the effective detector threshold. The goal for future PPC technology is to further reduce the threshold in these devices as close as possible to the 50 eVee (~0.2 keVnr) limit, while increasing the crystal mass by a factor of a few (the first C4 crystal weighs in at 1.3 kg, in contrast to the 0.4 kg CoGeNT PPC presently operating in Soudan). Sensitivity predictions can be found in [136].

## *Silicon CCDs*

The DAMIC experiment is using fully-depleted thick (320 micron) CCD's developed for astronomical survey instruments (SNAP/JDEM, DES) at LBNL. These sensors have been demonstrated in not-so-low background environments at Fermilab and have recently been deployed and are operating at SNOLAB. Initial data collected at SNOLAB indicates that backgrounds have not been reduced compared to shallow underground operation at Fermilab revealing considerable uranium and thorium backgrounds coming from the detector mounting structures. In addition to the common theme of radiometric background

---

reduction, the most significant challenge for this technology will be increasing mass to kg levels. These sensors have extremely low thresholds (few eV) and hence are sensitive to much lower WIMP mass scales than any other technology, and hence are complimentary to all other searches.

### *Bubble Chambers*

The COUPP/PICASSO experiments utilize bubble chambers to search for WIMP dark matter. Bubble chambers provide a "perfect" threshold cutoff that rejects low ionization loss events coming from electromagnetic backgrounds. COUPP is currently operating a 60 kg $CF_3I$ bubble chamber at SNOLAB.

The primary backgrounds for this technology are surface events coming from alpha-emitting materials in the quartz containment vessel and from neutrons penetrating the shielding. The technology is currently being scaled up to the ~100 kg level and a ~500 kg experiment has been proposed. Sensor challenges include vessel radiopurity and fluid purification systems needed to achieve the required background levels.

### *Solid Xenon*

The solid (crystalline) phase of xenon inherits most of the advantages of using liquid xenon as a detector target material for low energy particles; transparency, self-shielding, absence of intrinsic background, and ionization drift. In the solid phase, even more scintillation light yield (~60 keV) is reported compared to the liquid phase (~40 keV). Operation at sub-Kelvin temperature is natural for the solid phase, using superconducting sensors to read out photon, ionization, and phonon signals.

Most of the properties of solid xenon have been measured in a small volume or in thin film in the early 1970s through 1990s, but there were no systematic studies successfully carried out with large-scale solid xenon. In January 2010 Fermilab grew approximately a kilogram of transparent solid phase of xenon. Ongoing R&D at Fermilab is focused on scintillation light read out and electron drift in the solid xenon.

There are two remaining challenges for advanced detector development. (1) For solar axion searches, single-crystal xenon is favorable. The University of Enlargen (Germany) is planning to carry out the crystallography study of the solid xenon. (2) For neutrinoless double-beta decay searches, phonon read out from the solid xenon is favorable which requires millikelvin operation with superconducting phonon sensor technology. Characteristics of xenon at millikelvin temperatures have rarely been studied and deserves further R&D.

### *Materials and Assay*

The general approach to radioactive backgrounds is to reduce, shield, and finally discriminate against interactions with known types of ionizing radiation. The menagerie of direct detection experiments is a tribute to the clever approaches that have been developed to discriminate against electronic recoils from gamma and beta backgrounds. Typically, neutron backgrounds are the most insidious because they induce nuclear recoils identical to the expected WIMP signals and self-shielding against neutrons typically cuts a prohibitive amount of the active volume. Many of these material requirements are similar to other rare event searches in these energy ranges, for example low energy neutrino and neutrinoless double-decay experiments. However, the emphasis on neutron backgrounds is different than the emphasis on, say, 2 MeV gamma rays in germanium based neutrinoless double-decay experiments.

### *Materials*

Since neutrons backgrounds almost always limit an experiment, there is a special emphasis on removing uranium and thorium contamination since they can produce neutrons through spontaneous fission or ($\alpha$,n) depending on the matrix in which they are embedded. The amount of uranium and thorium contamination allowed in particular materials varies due to the location and matrix, but taking spontaneous fission as an inevitable process, a contamination of 1 mg of $^{238}$U results in a production of 1.2 neutrons per day. In context, the next generation dark matter experiments have background goals approaching <1 event/year in detectors with fiducial masses of ~1 ton.

Although there are a variety of dark matter experiments, a convenient dissection of the experiments is that they are made of shielding materials, detector materials, and electronics.

### *Shielding*

Traditionally, small detectors have used a mixture of lead to shield external gamma rays, and hydrogenated material, such as polyethylene, to shield external neutrons. As dark matter detectors have grown in size, more and more experiments are moving to water based shielding. Large (~3 meter thick) shields are common as these are thick enough to serve as both gamma and neutron shielding.

### *Detector Materials*

All dark matter detectors consist of some active element, germanium and xenon as examples, which are instrumented and contained in some sort of vessel. Some experiments require pressure vessels and others require vacuum vessels for containment. These vessels must satisfy leak and mechanical constraints. Materials currently used for containment include stainless steel, copper, and titanium. Generally, the radioactivity from these materials must be less than ~10 µBq/kg to meet the specifications of dark matter experiments.

### *Electronics*

The sensitive volumes are instrumented with various sensors and electronics. Many of the scintillator-based detectors utilize photomultiplier tubes (PMTs) and developing low radioactivity light detectors has been an active area of research. However, PMTs are often the dominant source of both neutron and gamma backgrounds in experiments based on the noble elements. Even with their high backgrounds, PMTs offer the low noise, high sensitivity, and large areas required by dark matter experiments, so they are the photodetector used in almost all scintillator based experiments. Other lower background devices such as large area avalanche photodiodes have been successfully deployed in other rare event searches, but the requirement for single photoelectron sensitivity currently precludes their use in dark matter detectors. Silicon photomultipliers are an attractive option, but their small areas and high dark rates are prohibitive.

Other, more basic, electronics components used in almost all dark matter experiments are potential sources of unwanted backgrounds including cables, circuit board substrates, resistors, and capacitors. Devices made from ceramics or with significant glass content typically contain significant uranium, thorium, and potassium contamination. A large amount of effort is spent by each collaboration to screen potential components and substrates to find sufficiently low background pieces. Often the radioactivity requirements are coupled with other unique requirements such as thermal conductivity.

### *Surface Contaminations*

It is not sufficient to screen materials for bulk contamination of radioactivity. Surface contamination, often from chemical deposition or radon plate out, is an important consideration for many rare event searches. This is especially true for surfaces that face the active detector material. Typical surface contaminants are due to chemical deposition and radon plate out, so these contaminants include isotopes that α-decay inducing worries about (α,n) neutron backgrounds from surfaces that do not have line of sight to the sensitive regions of detectors. The accumulation of radon daughters during material storage and handling is an important consideration for dark matter detector construction. These backgrounds can be difficult to assay at levels important for modern dark matter experiments since direct alpha, or low energy beta, screening is the most sensitive to radon daughters such as $^{210}$Pb.

## *Assay*

Assay of materials for radioactive isotopes is required to reduce risk and improve the scientific output of each dark matter experiment. Simulations of specific detector geometries is required to give contamination budgets for specific components. Then each candidate material is assayed to qualify a material or commercial product for a specific detector component. The most common methods of assay are through gamma counting with high purity germanium detectors (HPGe), mass spectroscopy, and neutron activation analysis (NAA). The requirements on sensitivity, sample size, and isotopic analysis will drive the decision on assay technique. High sensitivity HPGe analysis requires large samples, up to a few kilograms of material, whereas mass spectroscopy and NAA can achieve important radioactive contamination levels with ~1 g samples.

HPGe assay is a useful tool for material assay. The $^{238}$U and $^{232}$Th decay chains emit characteristic gamma rays, however the decay chains can be broken either through chemical means or at the radon break points, so gamma spectroscopy requires care at the stage of interpretation. Pure gamma emitting isotopes such as $^{40}$K are ideal candidates for HPGe assay.

Mass spectroscopy is a useful tool because it will measure exact isotopes of interest and extrapolations through decay chains is not required. Additionally, penetrating gamma rays are not required for elemental identification. Glow discharge mass spectroscopy (GDMS) is a useful tool because a variety of isotopes can be measured at the same time and sample preparation is minimal. However, GDMS has poor sensitivity, so is typically only useful for prescreening materials for more sensitive assay techniques. Inductively coupled plasma mass spectroscopy (ICPMS) is more sensitive to isotopes of interest and is an important tool for modern dark matter experimentalists. The most sensitive ICPMS requires dissolving samples for concentration, and special techniques are required to perform the most sensitive ICPMS on the large variety of materials required for dark matter detectors. ICPMS does suffer from important interferences that reduce the sensitivity for certain isotopes, for example $^{40}$K is difficult due to the argon gas used in ICPMS.

NAA is sensitive to a number of isotopes of interest for the dark matter community. Again this technique can probe specific isotopes of interest, such as $^{238}$U, without relying on decay chain assumptions. NAA requires careful sample preparation and specialized analysis. NAA, especially when combined with radiochemical separation, can result in some of the best minimum detection levels currently possible.

### *Custom Materials*

It has been shown that modern metallurgical techniques and special attention to polymerization precursors can reduce the radioactive contamination in materials. Collaborations interested in rare event searches have shown that custom materials can be developed when commercial materials do not meet specifications in terms of radioactivity.

For example, high-Z construction materials can be purified using electrochemical techniques to levels resulting in <1 µBq/kg $^{238}$U and $^{232}$Th activities. Not only can the radioactivity be reduced in these materials, but the mechanical properties can be controlled better than in typical high-throughput industrial techniques. This has been demonstrated for copper by the MAJORANA Collaboration, and could also be accomplished for other important metals such as gold, lead, nickel, tungsten and iridium as well as some metal alloys such as a number of copper alloys.

Custom low background electronics components such as capacitors, resistors, and cabling could be produced in sufficient quantities that the up-front development costs could be shared economically across future experiments reducing the need to constantly assay large numbers of these components.

*Summary*

Low background materials are important for dark matter research and their importance will only grow over the next decades. Insufficient care with seemingly insignificant detector components can drastically reduce the scientific impact of experiments [137], however sufficient care can produce experimental backgrounds within desired specifications [137]. The ability to source, assay, and keep materials free of radioactive contamination is crucial to reducing risk and improving scientific output from the efforts to directly detect WIMP dark matter interactions with terrestrial detectors.